\documentstyle[preprint,aps,pre]{revtex} 
\tighten                                 
\begin{document}                         
\draft                                   
\baselineskip=24pt
\newcommand{\beq}{\begin{equation}}
\newcommand{\eeq}{\end{equation}}
\newcommand{\noid}{\noindent}

\title{Growth, Percolation, and Correlations in
Disordered Fiber Networks}

\author{N. Provatas$^{1}$, M. Haataja$^{1,2}$, 
E. Sepp\"al\"a$^{3}$, S. Majaniemi$^{1}$,
J. {\AA}str\"om$^5$, 
M. Alava$^{3,4}$, and T. Ala-Nissila$^{1,2,6,*}$ 
}

\address{
$^1$Research Institute for Theoretical Physics, 
\\P.O. Box 9 (Siltavuorenpenger 20 C), 
FIN--00014 University of Helsinki, Helsinki, Finland
}

\address{
$^2$Laboratory of Physics, Tampere University of Technology,\\
    FIN--33101 Tampere, Finland}

\address{
$^3$Laboratory of Physics, Helsinki University of
Technology, \\
Otakaari 1 M, FIN--02150 Espoo, Finland} 

\address{
$^4$Department of Physics and Astronomy, Michigan State
University, East Lansing, Michigan 48824-1116
}

\address{
$^5$Department of Physics, University of Jyv\"askyl\"a,
P.O. Box 35, FIN--40351 Jyv\"askyl\"a, Finland
}

\address{
$^6$Department of Physics, Brown University,
Providence, Rhode Island 02912
}

\date{October 7, 1996}

\maketitle

\begin{abstract}
This paper studies growth, percolation, and 
correlations in disordered fiber networks. We start 
by introducing a 2D continuum
deposition model with effective
fiber-fiber interactions represented by a 
parameter $p$ which controls the degree of clustering. 
For $p=1$, the deposited network is uniformly random,
while for $p=0$ only a single connected cluster can grow.
For $p=0$, we first derive the
growth law for the average size of the cluster as well
as a formula for its mass density profile.
For $p>0$, we carry out extensive simulations on
fibers, and also needles and disks to study
the dependence of the percolation threshold on $p$.
We also derive a mean-field theory for the
threshold near $p=0$ and $p=1$ and find good qualitative
agreement with the simulations. 
The fiber networks produced by the model 
display nontrivial density correlations for $p<1$.
We study these by deriving an approximate expression for
the pair distribution function of the model that
reduces to the exactly known case of a
uniformly random network.  
We also show that the two-point mass density correlation
function of the model has a nontrivial form, and
discuss our results in view of recent experimental data on 
mass density
correlations in paper sheets. 

\end{abstract}


\bigskip
$^*$Corresponding author. E-mail address:
{\tt ala@phcu.helsinki.fi}.

\newpage
\section{Introduction}

There are many phenomena in nature that can be viewed
as deposition processes where various transport mechanisms 
bring particles to a surface.
These include many different 
examples such as
deposition of colloidal, polymer and fiber particles
\cite{Eva93,Pri95,Ono86,Dod94,Ala94,Nie95,Ryd92}.
Many deposition phenomena involve particles whose size is large
compared to the their mutual interaction range,
and so the main deposition mechanism is due to particle
exclusion.  Among the most studied
in this class are the Random and
Cooperative Sequential
Adsorption models \cite{Eva93,Pri95}.
There particles are deposited on a surface
and either stick or
are rejected according to certain exclusion rules,
with a maximum coverage (the ``jamming limit'')
less than unity.
These types of models should be contrasted to the case of
multilayer surface growth \cite{Eva93,Nie95,Ryd92}, where
the main focus is in the asymptotic behavior of the
growing surface in the continuum limit \cite{Bar95}.

A particularly interesting and challenging example involving
particle deposition can occur in
the case of colloidal suspensions.
For some such systems, the
interparticle repulsion is strong enough to prevent multilayer
growth \cite{Ono86}. However,
the existence of dispersion forces can cause the
particles to flocculate, or aggregate, and to precipitate out of
the suspension \cite{Ryd92,Mur95}. For larger particles or
clusters of particles, gravity often induces
sedimentation out of the suspension. Experiments reveal
that sedimentation produces nontrivial spatial structures
\cite{Kur96}. A full microscopic treatment of sedimentation
is a formidable task, however \cite{Sch95}.

An interesting practical deposition process is the formation 
of laboratory paper. Paper is manufactured from
a colloidal suspension, 
out of which fibers filtrate
onto a wire mesh, leaving behind a fiber 
network \cite{Dod94}. Recent
experimental measurements of mass density correlations in
laboratory made paper sheets reveal \cite{paperpaper} 
that the spatial distribution
of fibers may not be uniformly random.
In particular, for low mass 
density paper nontrivial power law
type of correlations of the local mass density
are found that may extend an order of
magnitude beyond the fiber length. 
This implies that there are nontrivial interactions present
during the formation process. A microscopic understanding of
this is currently missing. Because of this, phenomenological
deposition models may be useful in studying 
how various effective interactions influence the
mass density distribution of the deposit  
\cite{paperpaper}.

In addition to their practical applications, 
random deposition models
have been the topic of intense study in their own right. In particular, 
they have been extensively studied in the 
context of continuum percolation theory 
\cite{Pik74,Bal83,Rob83,Bal84,Rob84,Bug85,Bal87,Lar89,San90,Alo91,Van91,Dro94,Cho95,Rin95}. 
These models have included both uniformly random networks 
of various objects as well as some that include hard and soft-core 
interactions 
between the constituent particles. 
The quantity of central importance in these studies is the
percolation threshold or critical particle density 
which for permeable objects can be related to the excluded
volume of the particles \cite{Bal87}.
This quantity depends on the geometrical shape of the
deposited particles as well as on interactions between
them in a nontrivial way. 

Motivated by processes involving deposition,
sedimentation and
the theory of continuum percolation, we introduce 
and present a detailed study of the properties
of a simple 2D deposition model in this work. 
In the model, the tendency of the deposited particles to
form clusters is taken into account through a parameter 
$0\le p \le 1$.
For $p=1$, the deposition process is uniformly random
while for $p=0$, only a single connected cluster grows
from an initial seed fiber \cite{Pro95}.
We study the growth dynamics, percolation thresholds
and spatial correlations
of the structures formed by the model in detail, using
both analytic and numerical methods. In particular,
we discuss our results in view of the experimentally
observed mass density correlations in paper sheets \cite{paperpaper}.

The outline of this paper is as follows. We first introduce our
deposition model in Sec. II. In Sec. III, we examine the
growth properties of single clusters in the limit $p=0$.
After this we present both numerical and analytic results
for the percolation threshold of fibers and various other
objects as a function of $p$. Section V deals with the
theory of spatial correlations in networks formed by
the model; in particular we concentrate on the form of
the pair distribution and two point 
mass density correlation functions.
Finally, section VI contains our summary and
conclusions.

\section{The Model}

The model studied in this paper,
called the ``Flocculation Model'' is in part motivated
by the tendency of fibers to adhesively stick during
paper-making \cite{Dod94,Timo}. The model is a
continuum deposition model defined on a 2D surface. 
Spatially extended objects, such as 
disks and fibers with a finite area,
or widthless needles are individually
deposited by the following rules.
First, both the orientation of the object as well as its
spatial coordinates are chosen from a uniformly
random distribution. If a deposited object lands on another object
already on the surface, the attempt is always accepted.
However, if it lands on empty space the attempt is
accepted only with a given probability $p$. Thus, the
parameters that characterize the model are the
acceptance probability $p$, the dimensions
of the deposited objects, the linear dimension $L$ of the
surface, and the number of deposited objects kept $N$
(i.e. the number of accepted attempts).
In this work the deposited objects are mostly fibers
(rectangles of finite area and varying dimensions); 
however, in Sec. IV we also consider
needles (widthless sticks) as well as disks.
 
In the limit
$p=1$ the model reduces to the extensively studied case
of a uniformly random network 
\cite{Dod94,Pik74,Bal83,Rob83,Bal84,Rob84,Van91,Bou89}.
However, for $p<1$ it is clear that there are effective
interactions between the particles that tend to enhance
cluster formation within the deposit. In particular,
for the extreme case of $p=0$, only particles that always
touch each other are accepted, and a single, connected
cluster grows (assuming there is an initial seed
particle). We expect all the properties of the
growing network, including its percolation threshold
and spatial mass density correlations to depend on
$p$ in a nontrivial way.
We note that the definition of
the model here is different from
the extensively studied class of
Random Sequential Adsorption
models, where particle overlap is {\it not}
allowed \cite{Eva93}.

\section{Growth of individual fiber clusters}

First we examine the deposition of fibers in the
limit $p=0$ of the model \cite{Pro95} where only one, connected
fiber cluster grows from an initially deposited seed fiber. Fig. 1
shows a typical cluster, for $N=2000$  
fibers of length $\lambda=1$ and width $\omega=1/4$.
An individual cluster has an irregular, ragged shape for
which an average radius $R_i(N)$ can be defined. When
a statistical average is taken over all possible
clusters of size $N$, the average cluster is spherically
symmetric and one can define an average radius
$R(N) = \langle R_i(N) \rangle$ for which an expression
can be obtained by the following argument.
Consider the more general $d$ dimensional case of
a cluster of $N>>1$ identical 
fibers of linear dimension $a$ where  
additional $\Delta N $ fibers are deposited. 
Its radius grows by $\Delta R$ and area by $\Delta A$.
The change in the cluster area
$\Delta A \propto a^d  (R^{d-1}/ R^{d}) \Delta N \propto
R^{d-1} \Delta R$, which in the continuum limit
gives 

\begin{equation}
R(N)= B N^{1/(d+1)}, 
\label{clustergrowth}
\end{equation}
\noid
where $B$ is a constant proportional to $a$. In particular,  
for the two-dimensional case this means that the average
size of the cluster grows as $N^{1/3}$ as already shown in
Refs. \cite{Ast89,Pro95}.  

The radial mass density 
of the average fiber cluster was also examined in Refs.
\cite{Ast89,Pro95}. For $N>>1$ (typically $N > 500$)
required for the scaling relation Eq. (\ref{clustergrowth}), it
was shown numerically in Ref. \cite{Pro95} that all the 
density profiles scale onto one curve. The approximate form of this 
profile can be obtained by the following argument following Ref.
\cite{Ast89}. Let $\rho(r,N)$ denote the local mass density in a fiber
cluster of $N$ fibers, and $K$ the constant flux of mass per unit area onto
the cluster of radius $R(N)$. Assuming that the rate of change of $\rho$
is proportional to the flux and 
inversely proportional to the area of the cluster,
within the cluster radius, then we can write the equation for $\rho$ as 

\begin{equation}
\frac{ \partial \rho(r,N) }{\partial N} = \frac{K}{\pi R^2(N)}, \,\,0 < r
<R(N)\,\mbox{.}
\label{flux}
\end{equation}
\noid
For $r >R(N)$, we assume the flux to be zero. Using Eqns. (\ref{clustergrowth})
and (\ref{flux}) we obtain

\begin{equation}
\rho(r,N) = \frac{3K}{\pi B^3} R(N)(1-\frac{r}{R(N)}).
\label{density}
\end{equation}
\noid
Eq. (\ref{density}) is in good agreement with the numerically
obtained density profiles in \cite{Pro95}.
 
However, as already noted in Ref. \cite{Pro95}, at the edge of the average
cluster there is a systematic deviation from the linear decrease of
the density profile to a slower decay law. This indicates that the
edge of the cluster is rough, and that the roughness also grows
as a function of $N$. This can be characterized by calculating the
average width from the definition

\begin{equation}
w(N) = [\langle (R_i(N) - R(N))^2 \rangle]^{1/2}.
\end{equation}
\noid
Since the nonlinear edge of 
the density profile scales, too, we expect that $w(N)/R(N)=const.$
for $N>>1$. We have numerically simulated $w(N)$ in 2D and
indeed find that it grows proportional to $N^{1/3}$.

\section{Percolation properties of the model}

When enough fibers are deposited in a finite system, 
its edges become connected 
and percolation takes place in the model.
The properties of the corresponding 
{\it continuum percolation transition}
are of particular interest \cite{Bal87,Sta95}.
For the case of single cluster growth, the growth law of Eq. (1)
implies that $N \propto R^3$. Thus, within a finite system $L$
a single cluster spans the system only when $R \approx L$ which
means that the density of fibers $\eta = N/L^2 \propto L$, i.e.
it increases with the system size. This means that the
critical density $\eta_c$ for percolation becomes infinite
for $p=0$. However, for $p>0$ there is a finite probability 
of nucleation of multiple clusters and thus $\eta_c$ should be
finite. Indeed, continuum 
percolation of 2D rectangles \cite{Pik74,Bal84},
disks \cite{Pik74,Bal83,Bal84,Bug85,Bal87,San90,Alo91,Rin95} and
needles \cite{Pik74,Rob83,Bal84,Rob84,Van91,Bou89} 
among other
geometric objects has been extensively studied, and the
corresponding critical densities determined numerically.
In this section, we present results of numerical and analytic
calculations of the percolation properties of our model
for $0<p\le 1$ and compare the limit $p=1$ with the
existing studies.

\subsection{Numerical results}

We first present results of extensive numerical simulations
of the critical densities of percolation $\eta_c(p)$ as a 
function of $p$ for three different geometric objects:
needles of length $\lambda$, fibers (rectangles) of
length $\lambda$ and width $\omega$ ($\lambda > \omega$),
and disks of radius $r_d$. The details of the method we have
used are as follows. First, each 
system contains an inner box of length $L$ and 
an outer box $L+L'$, with free boundary conditions. 
It is important in the present case to
choose $L'$ to be large enough so that the average density
across the system is constant within the inner box \cite{box}.
The centers of the objects are distributed within 
the outer box. However, only objects partially or completely 
within the inner 
box are allowed to belong to the connected clusters. 
To keep track of the clusters
we employ a simple labeling technique \cite{Hos76}.

In analyzing the data we distinguish two different types
of percolation processes. In the one-sided case,
any two opposite edges of the inner box percolate while in
the two-sided case, all four edges must become connected.
In our final analysis, only one-sided percolation data
was used \cite{boundary}.
We obtained the critical density 
$\eta_{c}$ by the following procedure. We deposited
a fixed number of objects $N$ for a given system size $L$
and checked for percolation. 
By repeating this 100--1000 times
the spanning probability, i.e. the 
fraction of percolating ensembles were obtained. 
Repeating this procedure for various values of $N$, 
the whole curve of spanning probabilities was obtained for each
system size $L$, as a function of the density $\eta=N/L^2$.
The point where two such curves for any different system sizes
intersect, gives an approximate value for $\eta_c$. 
We obtained these points by fitting the curves to error
functions. The estimates thus obtained were
extrapolated and the final values of $\eta_c$ obtained 
using the standard Monte Carlo renormalization group 
method \cite{Rey80,San90}, with the smallest
system studied being the reference system. 
For completeness, we have also
evaluated the correlation length exponent $\nu$ for our
model from the MCRG procedure. We note that it is expected
to have the well-known value of $4/3$ as in lattice percolation 
\cite{Bal87,Sta95}.

\subsubsection{Needles}

Typical configurations generated by the model
are shown in Figs. 2 (a)-(c) for $N=2000$ needles for various
values of $p$ (1.0,0.8,0.6,0.4,0.2,0.05,0.01,0.005, and 0.001). 
The existence of well-defined clusters in the
network for the smallest value of $p=0.001$ is clearly 
visible. Employing the MCRG procedure for 
$1000$ ensembles and for
system sizes $L=6,20,30,40,60$ for $0.2 \leq p \leq 1.0$ and
$L=6,20,40,60$ for $0.005 \leq p \leq 0.05$, we obtain
the values for $\eta_c(p)$ as displayed in Fig. 3. 
The curve displays interesting behavior in the two limits
$p \rightarrow 0$ and $p \rightarrow 1$. The expected divergence of 
$\eta_c(p)$ in the limit $p \rightarrow 0$ is clearly visible,
although for smallest $p$ our results here and
below are only qualitative
due to strongly increasing finite size effects. 
Moreover, $\eta_c(p)$ approaches the limit $p \rightarrow 1$ 
approximately linearly with a minimum at $p \approx 0.2$.
These results together with the minimum in the curve
visible around $p\approx 0.2$ will be further discussed in
Sec. IV B.

Our best estimate for $\eta_c(1)=5.64 \pm 0.02$ agrees 
rather well
with other numerical studies reported in the literature for
the uniformly random case
(see, {\it e.g.}, Ref. \cite{Van91}).
It also agrees with the estimate
of Ref. \cite{Bal84} which gives a value of
5.61.
Figure 4 shows the effective
correlation length exponent $\nu(p)$ as evaluated
from the MCRG procedure. We note that $\nu(p)$ is consistent with 
$\nu = 4/3$ in the limit $p \rightarrow 1$. However, for 
$p\rightarrow 0$ the effective $\nu$ 
increases. In Sec. V we show that this
is due to the increasing range of spatial correlations
in the case where individual clusters become large.
This induces strong crossover effects towards
long-range correlated percolation type of 
behavior \cite{Wei84}.

\subsubsection{Fibers}

In the case of fibers, we have used objects with
an aspect ratio of 
$\lambda / \omega = 4 $. Typical networks generated
are shown in Figs. 5 (a)-(c) for various values of $p$
(1.0,0.9,0.8,0.7,0.5,0.4,0.3,0.2,0.1,0.05,0.01,0.005, and 0.001).
Following the procedure employed for
needles for system sizes $L=20, 40, 80$, we obtain the 
$\eta_c(p)$ curve as shown in Fig. 6. 
These results have been
obtained by averaging over a $100$ ensembles. Again, we see
the expected divergence of $\eta_c(p)$ 
as $p$ goes to zero. However, in the other limit
$p \rightarrow 1$, $\eta_c(p)$ is almost
constant within the error bars. 
Our best estimate for the $p=1$ limit
is $\eta_c(1)=2.73 \pm 0.2$. 
To our knowledge, there are no other numerical studies
of $\eta_c(1)$ for the present aspect ratio.
However, $\eta_c(1)$ can be approximately
determined by using the excluded volume arguments
of Ref. \cite{Bal84}, where $\eta_c(1) \langle
A \rangle = const. \approx 3.57$, and the
excluded volume $\langle A \rangle =1.379$
\cite{excluded}. This gives results within
about 10 \% agreement with our numerical data.
Finally, the correlation length exponent $\nu(p)$ 
is shown in Fig. 7. Again, as $p \rightarrow 1$, $\nu(p)$ 
is consistent with $\nu = 4/3$. Moreover, 
the effective $\nu(p)$ again increases as $p \rightarrow 0$.

\subsubsection{Disks}

For completeness, we have also studied the case of isotropic disks. 
Typical networks of $N=1000$ disks of radius $r_d=0.5$ for various values
of $p$ 
(1.0,0.9,0.8,0.7,0.6,0.5,0.45,0.4,
0.35,0.3,0.25,0.2,0.15,0.1,0.05,0.02,0.01,0.005, and 0.001)
are shown in Figs. 8 (a)-(c). Following the procedure employed for
needles for system sizes $L=4,24,32,44,64$ and averaging
over $800$ ensembles for disks of radius $r_d=1$, we obtain the $\eta_c(p)$ 
curve as shown in Fig. 9 \cite{onesided}. Again, $\eta_c(p)$ diverges as 
$p \rightarrow 0$. However, for about $p>0.3$,
$\eta_c(p) \approx const.$ within the error bars. 
We note that our best estimate for $\eta_c(1)=0.374 \pm 0.016$ agrees
within the errors with other numerical studies reported 
in the literature (see, {\it e.g.}, Ref. \cite{San90}).
The correlation length exponent $\nu(p)$
is shown in Fig. 10. Again, as $p \rightarrow 1$, $\nu(p)$ 
is consistent with $\nu = 4/3$. Also, 
the effective $\nu(p)$ increases as $p \rightarrow 0$.

\subsection{Mean-field theory for percolation thresholds}
\label{mf for pth}

We can {\it qualitatively} understand the behavior of the percolation
thresholds for the two limits where $p\rightarrow 0$ and
$p\rightarrow 1$ using mean-field theory.  
First, let us discuss the case where $p\rightarrow 0$.
To this end, let us define 
$q$ as the probability of a given object intersecting any other object
in a uniformly random network ($p=1$). This quantity
depends on the dimensions of the object, and the system
size $L$. For example, in the case of fibers 
\cite{Kal60}

\begin{equation}
q=\frac{2(\lambda+\omega)^2}{\pi L^2}+\frac{2\lambda\omega}{L^2}.
\label{q}
\end{equation}
\noid
For needles, one can just set $\omega=0$. Moreover, for disks
a simple calculation gives $q=9\pi r_d^2/L^2$.

Next, consider the actual limit of
$p\rightarrow 0$. Suppose we begin with an initially 
empty system and drop one object, and then drop another. 
The probability that it sticks onto the 2D plane but {\it not} 
on the first one is $P_s^{(1)}=(1-q)p$. 
The probability that it {\it does} stick on top of the first one 
is $Q_s^{(1)}=q$. Continuing this process for $n$ particles we find 
correspondingly that 
for the $n$th particle, $P_s^{(n)}=(1-q)^{n-1}p$, while 
$Q_s^{(n)}=1-(1-q)^{n-1}$. Note that for sufficiently
small $q$, we assume that the initially objects tend to land in 
a uniformly random manner.  For $n$ greater than some $n=n_c$ 
there comes a point where every additional object will
encounter as much occupied
as empty space. From this point onwards, 
additional objects will essentially tend to build up clusters around the
initial $n_c$ seed objects.
By this reasoning, the 
mean number of clusters $n_c$ can be approximately obtained
from the solution of 

\begin{equation}
(1-q)^{n_c-1} p= 1-(1-q)^{n_c-1}
\label{nuc1}
\end{equation}
\noid
which gives

\begin{equation}
n_c = 1-\frac{\log(1+p)}{\log(1-q)}.
\label{nuc2}
\end{equation}
\noid
We note that this equation is expected to {\it underestimate}
the true $n_c$ due to the fact that due to clustering for $p<1$, 
the effective excluded area
of the objects will be smaller than that given by the bare
$q$ in Eq. (\ref{q}). 
Let us assume that the growth of any one of the $n_c$ clusters follows
Eq. (1)

\begin{equation}
R_c=B N_c^{1/3}, 
\label{growth}
\end{equation}
\noid
where $N_c$ is the number of particles in each cluster. Note that 

\begin{equation}
N=n_c N_c, 
\label{ntot}
\end{equation}
\noid
where $N$ is the total number of objects in the
network. On the average, the system must percolate
when the clusters begin to overlap, {\it i.e.}, when

\begin{equation}
\sqrt{\frac{1}{\rho}} \approx 2R,
\label{criterion}
\end{equation}
\noid
where $\rho$ is the density of clusters, $\rho= n_c/L^2$.

Substituting Eq. (\ref{nuc2}) into the expression for $\rho$ and using 
Eq. (\ref{criterion}) we obtain, 

\begin{equation}
2 B N_c^{1/3} =L \left( 1-\frac{\log(1+p)}{\log(1-q)} \right)^{-1/2}.
\label{inter}
\end{equation}
\noid
Using Eq. (\ref{ntot}), we obtain 

\begin{equation}
\eta_{c,L} = \frac{L}{8 B^3} \left( 1-\frac{\log(1+p)}{\log(1-q)}
\right)^{-1/2}.
\label{finitesize}
\end{equation}
\noid
Equation (\ref{finitesize}) gives us the mean-field percolation
threshold defined as
$\eta_{c} \equiv \lim_{L \rightarrow \infty} \eta_{c,L}$.
In particular, for the case of fibers we obtain 

\begin{equation}
\eta_{c}=\sqrt{\frac{(\lambda+\omega)^2+\lambda \omega \pi}{32 \pi B^6}} 
\log^{-1/2}(1+p).
\label{infinitesize}
\end{equation}
\noid
This result is expected to become more and more accurate
in the limit $p \rightarrow 0$, where it becomes 

\begin{equation}
\eta_{c} = \sqrt{\frac{(\lambda+\omega)^2+\lambda \omega \pi}{32 \pi B^6}}
\sqrt{ \frac{1}{p}}.
\label{smallp}
\end{equation}

The result that $\eta_{c} \propto \sqrt{1/p}$ has  also been obtained in a 
somewhat different manner in Ref. \cite{Ast89}, in accordance
with our results. This universal inverse 
square root divergence on $p$ is
in qualitative agreement with the numerical data presented in Figs.
3, 6 and 9. However, due to the numerical difficulties in
determining $\eta_c$ accurately in this region, we have not
been able to quantitatively verify Eq. (\ref{smallp}).
It is interesting to note that in the small $p$ limit, where
precolation occurs through the growth of well-defined individual
clusters, our model is somewhat similar to the cooperative
lattice filling model where cluster growth leads to percolation
\cite{San88}. However, in this case there is no divergence of
$\eta_c$ since overlap is not allowed.

Next we consider the other limit of $p \rightarrow 1$.
Let us again start with a purely uniformly random case ($p=1$) 
at its percolation
threshold. The probability of any two particles intersecting 
is given by $q$ as before. 
Take an arbitrary particle from the network
of $N$ particles. The probability $P_s^{(N)}$ is now given by

\begin{equation} \label{pint}
P_{s}^{(N)}=(1-q)^{N-1}.
\end{equation}
\noid
Therefore, the number of non-intersecting objects will be,
on the average,
 
\begin{equation}
N_{s} = P_{s}^{(N)} N.
\end{equation}
\noid
The main {\it qualitative} idea in the derivation here
is the following. The effect of 
$p < 1$ is to reduce
the number of non-intersecting objects due to
clustering, but the backbone of the percolation cluster
itself can be assumed to stay constant for $p$ not too small.
This already implies that the percolation threshold 
decreases as $p$ decreases.

Let us next quantify this argument.
Assume that 
the number single particles removed that do not belong to
the percolation cluster, is given by
\noindent

\begin{equation}
N_{r}=(1-p) \,N_{s}.
\end{equation}
\noid
Therefore, the number of particles left at the
percolation threshold for such $p$
can be written as

\begin{eqnarray}
N(p) = N(1) -(1-p) \,N_{s} \nonumber \\
 = N(1) - (1-p) \, P_{s}^{(N)} \, N(1),
\end{eqnarray}
\noid
where $N(1)$ is the total number of particles within the
uniformly random reference system with $p=1$. 
For the critical density we thus obtain

\begin{eqnarray}
\eta_{c,L}(p) = \eta_{c,L}(1) - P_{s}^{(N)} \eta_{c,L}(1)
 + p P_{s}^{(N)} \eta_{c,L}(1) \nonumber \\
 \equiv Ap + B,
\end{eqnarray}
\noid
Therefore, we can conclude that $\eta_{c,L}(p)$ increases
linearly with $p$. This is also visible in our numerical data
for the critical densities, at least for needles and
fibers. To examine the theory 
quantitatively, we can show that for the case
of fibers, in the limit
$L \rightarrow \infty$, 

\begin{equation}
A=\eta_c(1)
\exp[-\eta_c(1)(\frac{2(\lambda+\omega)^2}{\pi}+2\lambda\omega)].
\end{equation}

We have estimated $A$ from the numerical data
for needles, and find that the numerical and theoretical
values agree within about $50$ \%
\cite{clusters}. The theoretical slope for
disks is about an order of magnitude smaller than for needles, and
thus it is not surprising that 
it is not discernible in our data of Fig. 9
within the error bars. Also, for the case of fibers the error
bars are too large to see the slope. 

The analytic derivations of this section can be used to
explain the minimum in the $\eta_c(p)$ vs.
$p$ curves. Since in the limit $p
\rightarrow 0$, $\eta_c(p)$ must diverge, whereas for
$p$ slightly below unity $\eta_c(p)$ must linearly
decrease, it follows that a minimum must exist for some
$0 < p < 1$. Physically, the existence of this minimum
can be explained by a
competition between two mechanisms in the model.
First, 
single cluster growth that dominates for small $p$
is rather an inefficient 
way of forming the percolating cluster. On the other
hand, uniform random filling of the lattice is
inefficient, too, since there are many particles
that do not belong to the percolating cluster.
Thus, there is an ``optimum'' way of forming
the cluster at some intermediate value of $p$, where
the critical density is minimal. 
 
\section{Pair distribution and the two-point mass density correlation
function }

In this section we begin by writing down 
a general expression for the two-point 
mass density correlation function for a random network of objects
in which there is
some effective particle-particle interaction. 
We then generalize this expression for the case of fiber networks.
This expression involves the pair distribution function for the
centers of the objects. This function has previously been
derived exactly for the case uniformly random network of 
objects \cite{Dod94,Gho51}.
We generalize this derivation for our model,
and compare the results with numerical simulations. We also
examine in detail the form of the two-point mass density correlation
function for the model.

\subsection{A general expression for the two-point mass
density correlation
function}

The two-point density correlation function
is generally defined as \cite{Hau74}
 
\begin{equation}
G(r) =\lim_{A \rightarrow \infty} f^{-1} (\frac{| \hat{\rho}(k) |^2}{A}),
\label{cordef}
\end{equation}
\noid
where $f$ is the 2D Fourier transform operator, 
$\hat{\rho}(k)$ is the Fourier transform of the mass density distribution 
and $A=L^2$ is the area. This distribution can be written as 

\begin{equation}
\rho(\vec x) = \sum_{i=1}^{N} \mu(\vec x,\vec x_i,\theta_i)
\label{massdef}
\end{equation}
\noid
where $\mu(\vec x,\vec x_i,\theta_i)$ is equal to $\sigma$ over the 
surface of particle $i$ which is located at the point
$\vec x_i$ with orientation $\theta_i$, and zero outside of
it. Here $\sigma$ is the mass density of each particle.
Writing 
$\hat{\rho}(k)=\int_A  \rho(\vec x) e^{-2 \pi i \vec k \cdot 
\vec x}$, and substituting 
into Eq. (\ref{cordef}) we obtain after some algebra
the general expression \cite{corref}

\begin{equation}
\hat{n}(k) = \eta \langle |I(k)|^2\rangle + \langle|I(k)| \rangle^2 
\lim_{A \rightarrow \infty} \frac{N(N-1)}{2 A} \left( 
\int_{A} \int_{A} e^{-2 \pi i \vec k \cdot (\vec x_i -\vec x_j)} \tilde{g}
(|\vec x_i - \vec x_j|) d^2 x_i d^2 x_j \right),
\label{corr1}
\end{equation}
\noid
where  
$\hat{n}(k) \equiv \lim_{A \rightarrow \infty} 
|\hat{\rho}(k) |^2 / (A \sigma^2)$,
and $\eta$ is the number density of the particles.
This gives the correlation function as

\begin{equation}
G(r) = f^{-1}(\sigma^2 \hat{n}(k)).
\label{gdef}
\end{equation}

In the equations above, $\langle|I(k)|^2\rangle$ 
denotes the square of the 
Fourier transform of a single object
averaged over all orientations. 
The second term in Eq. (\ref{corr1}) contains
the center-center pair distribution function
$\tilde{g}(|\vec x_i -\vec x_j|)$ \cite{corref} that describes
the effective interparticle interactions in the network. It 
gives the probability per unit area squared
of particle center $i$ occupying position $\vec x_i$ 
and that of another particle center $j$ 
occupying position $\vec x_j$.  
Making a change of variables and carrying 
out one of the integrals in Eq.~(\ref{corr1}) gives

\begin{equation}
\hat{n}(k)= \eta \langle |I(k)|^2\rangle + 
\eta^2 \langle |I(k)|\rangle^2 
\lim_{A \rightarrow \infty} \left(
\int_{0}^{\sqrt{2} L} J_0(2 \pi k r ) [A \Omega(r)] dr
\right),
\label{corr2}
\end{equation}
\noid
where $J_0(x)$ is the Bessel function of order zero, and 
we define $A \Omega(r)
\equiv  2\pi r  \tilde{g}(r)$. 
Equation (\ref{corr2}) has two components. The 
first describes the self-correlation 
of single particles, while the second 
describes the cross-correlations, through the
function $\Omega(r)$. The function $\Omega$ is
defined as

\begin{equation}
\Omega(r) dr  = 
\Bigl \langle \frac{\mbox{\# of pairs of centers in a shell 
({\it r, r}+{\it dr})}}
{\mbox{total \# of pairs of centers in system}} \Bigr \rangle,
\label{grdef}
\end{equation}
\noid
where the averaging is over
all configurations. The function $\Omega(r)$ in our final expression
Eq. (\ref{corr2}) contains the effects of particle interactions,
while $I(k)$ contains the information about the geometry of the
deposited objects. In the special case of fibers, $I(k)$ has been  
derived exactly in Ref. \cite{Hau74} giving 

\begin{equation}
\langle|I(k)|^2\rangle=(\lambda \omega)^2 
\left( \frac{1}{2 \pi} \int_{0}^{2 \pi} \mbox{sinc}[\pi k \lambda \cos(t)] 
\mbox{sinc}[\pi k \omega \sin(t)] dt \right)^2
\end{equation}
\noid
and

\begin{equation}
\langle |I(k)|\rangle^2= (\lambda \omega)^2 
\frac{1}{2 \pi} \int_{0}^{2 \pi} \mbox{sinc}^2 [\pi k \lambda \cos(t)] 
\mbox{sinc}^2[\pi k \omega \sin(t)] dt ,
\end{equation}
\noid
where $\mbox{sinc}(x)=\sin(x)/x$. In the previous
studies, the second term 
in Eq. (\ref{corr2}) has been neglected for uniformly
random fiber networks. However, 
for our model, where effective interactions arise for $p<1$ the 
second term gives rise to nontrivial correlations that must
be taken into account.  

\subsection{The pair distribution function}

An exact expression for $\Omega(r)$ in the 
case of uniformly random networks
has previously been derived by Ghosh \cite{Dod94,Gho51}.
Below, we will generalize this derivation to obtain
an approximate formula for our model that
is valid for small values of $p$.
To begin with, 
let $\Delta N_{\mbox\scriptsize CM}(\vec{x}_i)$ be the number of centers of
masses within the area element $\Delta A(\vec{x}_i)$
around the position vector $\vec{x}_i$. The number of pairs included in two
different area elements is then given by the product
$\Delta N_{\mbox\scriptsize CM}
(\vec{x}_i)\Delta N_{\mbox\scriptsize CM}(\vec{x}_j)$. Further, the total
number of pairs separated by vector $\vec{x}$ in a given 
configuration is

\beq
\sum_{\vec{x}_0} \Delta N_{\mbox\scriptsize CM}
(\vec{x}_0)\Delta N_{\mbox\scriptsize CM}(\vec{x}_0 + \vec{x}).
\eeq
\noid
Dividing this by $N(N-1)/2$ we obtain the probability to find a pair
of centers of mass separated by the vector $\vec{x}$ in a given
configuration.
Multiplying and dividing by area elements yields

\begin{equation}
\Omega(r) dr = \frac{2}{N(N-1)} \Bigl \langle \sum_{\vec{x}_0}
\frac{\Delta
N_{\mbox\scriptsize CM}
(\vec{x}_0)}{\Delta A(\vec{x}_0)} \frac{\Delta N_{\mbox\scriptsize
CM}(\vec{x}_0 + \vec{x})}{{\Delta A(\vec{x}_0 + \vec{x})}}\,\Delta A(
\vec{x}_0)\Delta A(\vec{x}_0 + \vec{x}) \Bigr \rangle,
\end{equation}
\noid
where we have assumed that the system is isotropic and thus
$\Omega$ depends on $r\equiv \vert \vec{x}\vert$ only.
The brackets denote configuration averaging.
Taking the continuum limit where $\eta_{\mbox\scriptsize CM}
= \lim_{\Delta A \rightarrow 0} (\Delta N_{\mbox\scriptsize CM}/
\Delta A)$ we obtain

\begin{equation}
\Omega(r) dr =
\frac{2}{N(N-1)} \langle \int_A d^2x_0 \eta_{\mbox\scriptsize
CM}(\vec{x}_0) \eta_{\mbox\scriptsize CM}(\vec{x}_0 + \vec{x})
\Delta A(\vec{x}_0 + \vec{x})\rangle. 
\label{bas}
\end{equation}
\noid
Hence,

\beq
\Omega(r) dr  =  \frac{2}{N(N-1)}
\int_A d^2x_0 G_{\mbox\scriptsize CM}(r_0, r) \Delta A(r_0 +
r)
\label{ty1},
\eeq
\noid
where

\beq
G_{\mbox\scriptsize CM}(r_0, r) \equiv \langle
\eta_{\mbox\scriptsize CM}(
\vec{x}_0) \eta_{\mbox\scriptsize CM}(\vec{x}_0 + \vec{x}) \rangle.
\label{tr}
\eeq

It is easy to see that for a uniformly random set of points
with translational invariance 
$G_{\mbox\scriptsize CM}(r) = N(N-1)/(2A^2)
=const.$ (no double counting of pairs),
which leads to $A \Omega(r)=
\Omega_h(r,L)$ with
$\Omega_h$ being the previously derived 
exact pair distribution function for a uniform random
network \cite{Dod94,Gho51}. For large system size ($A \to \infty$),
$A\Omega(r) \to 2 \pi r$. When this is substituted into 
Eq. (\ref{corr2}), a constant term arises in the correlation
function since there are no particle-particle interactions in this
special case. 

In the case of $p<1$ where clustering occurs in the 
model, the pair
distribution function $\Omega(r)$ 
gives rise to nontrivial correlations. In this
case $\Omega(r)$ can still be derived approximately
using mean field arguments. A straightforward but lengthy
calculation gives the result (see the 
Appendix for details)

\beq
\Omega(r)  =  \omega_1(N,p) g(r) + \omega_2(N,p)
\Omega_h(r,L),
\label{main}
\eeq
\noid
where

\beq
g(r) = \left\{\begin{array}{ll}
(30 r/42R^5)(3r^3 - 6Rr^2 + 4R^3)\, ,  &
\mbox{for $0 \leq r \leq R$;} \\ 
(30 r/42R^5)(2R - r)^3\, , &
\mbox{for $R \leq r \leq 2R$;}\\
0\, , & \mbox{for $2R \leq r$.}
\end{array}\right.
\label{gog}
\eeq
\noid
The coefficients are defined as $\omega_1(N,p) \equiv n_c
N_c(N_c - 1)/[N(N - 1)]$, where $n_c$ and $N_c$ are defined
in Sec. IV, and $\omega_2(N,p) = 1 - \omega_1(N,p)$. 
In Eq. (\ref{main}) $\Omega_h(r,L)$ denotes 
the pair distribution function for a uniformly random system
that is given by \cite{Dod94,Gho51}

\beq
\Omega_h(r,L) = \left\{\begin{array}{ll}
(4r/L^4)[\pi L^2/2 - 2rL + r^2/2]\ , &
\mbox{for $0 \leq r \leq L$;} \\
(4r/L^4) [L^2 \left(
\arcsin(L/r) - \arccos(L/r)\right) & \\
+ 2L\sqrt{r^2 - L^2} -\frac{1}{2}(r^2 + 2L^2)]\ ,
& \mbox{for $L \leq r \leq \sqrt{2}L$\,\, .}
\end{array}\right.
\label{unif}
\eeq

As discussed in more detail in the Appendix, Eq. (\ref{main})
is only approximately valid, although it does reduce to the
exact limit $\Omega_h(r,L)$ for $p=1$.
It has been derived under the assumption that the deposited
objects form clusters that grow independently
following the scaling relations derived in Sec. III.
This approximation is best justified in the 
$p \rightarrow 0$ limit. 

In Figs. 11(a) and (b) we illustrate 
several numerically 
obtained curves for $\Omega(r)$ for $p=0.001$ and 0.01, respectively,
in the case of fiber networks.
There are two common features in the figures.
First, for small values of $N$ well below percolation,
$\Omega(r)$ develops a sharp peak near the origin. This
peak is due to the 
cluster structure in the growing network.
Moreover, its position is proportional to the average size
of the clusters as can be seen from Eq. (\ref{gog}).  
In the opposite limit of $N \rightarrow \infty$, 
$\Omega(r)$ approaches $\Omega_h(r,L)$ as expected, since
the effect of $p$ becomes gradually less 
important after the network has percolated.
In the intermediate regions, a combination of the
two effects is seen and a smooth transformation occurs
between the two regimes. We note that when $p$ increases,
the height of the first peak decreases and its
position shifts to smaller values of $r$ indicating that
clusters become smaller.

To quantitatively test the accuracy of our analytic formula,
we compared the numerically obtained function $\Omega(r)$
with our analytic result of Eq. (\ref{gog}).
In Figs. 12 (a) through (c) we show the results of comparison
for $p=0.001$, with varying $N$.
In evaluating the analytic formula
two quantities, namely the average
cluster radius $R$ and the 
number of clusters $n_c$ are needed.
The latter can be evaluated from Eq. (\ref{nuc2})
for any $p$. However, since this equation
underestimates the true
number of clusters in the system, e.g. for $p=0.001$
it gives $n_c \approx 1.2$, we have allowed $n_c$
to be a fitting parameter. 
The quantity $R$, on the other hand, is fixed using the scaling relation
$R=BN_c^{1/3}$ where $N_c$ is the number of fibers in
each cluster. This can be evaluated from the identity $N=n_cN_c$
(see Sec. IV B). The only parameter left is the
prefactor $B$ which we fix to be $B=0.8$. This value
has been extracted from numerical data for single cluster growth.
As seen in Figs. 12 (a)-(c), setting $n_c=4.5$ gives very good
agreement between the theory and the simulations. This 
agreement also lends indirect support to the inverse square root
divergence of the percolation threshold on $p$ derived in Sec.
IV B.

We should note that the derivation of the
analytic formula is expected to be valid at most 
up to the point of percolation and for small values of $p$, since
otherwise the whole picture of growing independent clusters
breaks down. We have not carried out a systematic study
of the range of validity of the theory for larger values
of $p$, however.

\subsection{The two-point mass density correlation function}

Next, we examine the two-point mass density correlation function 
for the model. In particular, we examine the case of
fiber networks with an aspect ratio of 
$\lambda/\omega =20/1$, which is rather close to the needle limit.
We also discuss the effect of 
different object geometries on the correlation function.
It can be obtained from our formalism
by substituting the numerically 
evaluated $\Omega(r)$ into Eq. (\ref{corr2}), and using
Eq. (\ref{gdef}). However, for simplicity 
we have calculated the correlation function 
directly from simulations on a lattice \cite{fnote}.
Defining $\tilde{G}(\vec{x}) \equiv
\langle[ m(\vec{x}')-\langle m \rangle][ 
m(\vec{x}'+\vec{x})- \langle m \rangle] 
\rangle$ to
be the mass density-density {\it fluctuation} correlation
function, we find that its form is
well approximated by

\begin{equation}
\tilde{G}(r) \sim r^{-\alpha(N,p)}, \,\,\,\,{\rm for} \, 
0 < r < \Lambda(N,p),
\label{powerlaw}
\end{equation}
\noid
where $\Lambda$ is an effective cutoff for the power law form.
In the case of $p=1$, we obtain 
$\alpha \approx 1$, and $\Lambda \approx \lambda/2$. For lower
values of $p$, however, $\tilde{G}(r)$ displays interesting
behavior, as illustrated in Fig. 13, where we show 
$\tilde{G}(r)$ on a logarithmic scale for $p=0.001$. By 
examining the curves it can be inferred that $\alpha(N,p)$ goes 
through a minimum as $N$ increases. Moreover, $\Lambda$ attains 
a maximum where $\alpha$ is minimum. 

Figure 14(a) shows the effective $\alpha(N,p)$
vs. $N$ for the model, for different values of $p$.  
The behaviour of $\alpha$ can be understood as a 
competition between individual fiber clusters and uniformly random fibers, 
both of which coexist for $0 < p < 1$. The approach 
of the correlation exponent toward $\alpha=1$ as $N \rightarrow \infty$ 
is caused because in this limit the effect of 
$p$ gradually becomes unimportant as discussed in Sec. V B. 
On the other hand, 
the initial decrease of the effective $\alpha(N,p)$ for small values
of $N$ (as well as the increase of 
the range of the corresponding power law regime) 
is due to the growth of essentially
independent fiber clusters. 

A better understanding of the low $N$ behaviour of 
$\alpha(N,p)$, 
as well as the increase in the power law regime, can be understood
by considering the particular case of $p=0$. In this limit, only a 
single cluster emerges (see Fig. 1).
We find that $\tilde{G}(r)$ as averaged over an ensemble of
clusters of size $N$
is also well approximated by a power law form. 
Figure 15 shows $\alpha(N,p=0)$ 
vs. $N$ for this case. For the case of a large enough $N$
(which is about 300--400 for the fibers here), 
$\alpha$ saturates to a fixed value of about $0.05$. 
Moreover, we also find
that the range of the power law is proportional 
to the radius of the cluster, 
with a constant of proportionality of about 
$0.7$. We note that $\alpha$ reaches its saturated value
approximately at the same value of $N$ where the scaling laws
of single cluster growth in Sec. III start to become valid.
 
We have also examined the correlation function
for the fiber aspect ratio
of $\lambda / \omega=20/5$ 
that is away from the needle limit, and
for isotropic disks. In the case of such fiber or disk
networks, we find that for values of $N$ and $p$
comparable to those of the needle-like case,  
the correlation 
function is not as well approximated
by a simple power law behavior.
However, as expected the power law behavior does
eventually emerge for small enough values of $p$. 
The same conclusion also applies to single cluster
correlations where larger values of $N$ are needed.
Thus, we can conclude that the needle-like anisotropy
enhances the emergence of power law type of behavior in
the system.

It is instructive to consider a random cluster network 
composed of a uniformly random distribution of fiber clusters, each 
of $N$ fibers, in the needle-like limit.
Following the method of Sec. V, 
it is straightforward to show that the 
correlation function $\tilde{G}(r)$ 
for such a network is given by Eq. (\ref{corr2}), 
where $\langle|I|^2\rangle$ 
now represents the square of the Fourier transform of a 
{\it single} cluster. 
In this case $\tilde{G}(r)$ is exactly proportional
to the correlation function of single clusters discussed
above, and the same power law form emerges, with
its range proportional to the cluster size.
However, we note that as discussed in Ref. \cite{Pro95},
even longer range power law type of correlations for a
network of fiber clusters can be obtained by
assuming that there is an effective cluster-cluster
interaction that gives rise to such correlations
beyond the cluster size. 

It is interesting to examine our results for the 
areal mass density
fluctuation correlation function 
with recent measurement of this quantity 
for laboratory made paper \cite{paperpaper}.
It has been shown that $\tilde{G}(r)$ 
for these paper sheets is well approximated by 
a power law form whose exponent is independent of the 
mass density
of paper, and is $\alpha\approx 0.37$. 
Moreover, the range of this power law is rather short 
for high basis weight 
paper, i.e. of the order of the fiber length,
whereas it becomes longer ranged for low 
mass density paper extending even
up to about 14 times the fiber length. 
It is interesting to note that the simple deposition model
presented here can give similar values of $\alpha$ for
values of $N$ corresponding to real paper, by suitably
adjusting $p$ (see Fig. 14). However, the range of the
power law in this case does not vary with mass density in the
same way as in real paper. We can conclude from the results 
of this work that there are at least two possibilities of
obtaining longer range correlations such as in low mass density
paper. The first is that paper comprises small clusters
between which there is an effective interaction giving rise
to such correlations as explained in Ref. \cite{Pro95}. 
The second is that paper can be described
by deposition of larger, independent clusters whose
internal correlations are of power law type, with $\alpha
\approx 0.37$ \cite{paperpaper}. However, the true
microscopic modeling of paper-making and unraveling
of the reasons behind spatial correlations is a formidable
theoretical problem.

\section{Summary and conclusions}

In this paper we have studied growth, percolation and correlation
properties of disordered networks of fibers and other extended
objects. In particular, we have introduced a
deposition model that introduces an effective sticking
of particles upon deposition.
The key parameter of the model is the acceptance probability $p$
which describes the effective interactions between particles, thereby 
controlling the tendency to form clusters. This choice of rule
is motivated by interactions that may be present in suspension during
the formation phase of paper.

By varying $p$ the model interpolates 
between the growth of uniformly random networks ($p=1$) 
and single cluster growth ($p=0$). 
We started by studying the growth of single clusters, and
the form of the density profiles for such clusters. For
both quantities, analytic results have been derived and
verified by numerical simulations.
Following this, we examined the percolation 
threshold $\eta_c(p)$ of
needles, fibers, and disks in detail by performing extensive
numerical calculations as a function of $p$. In the limit
$p=1$, our results reduce to those previously obtained
for needles and disks. In the case of rectangles, however,
we were unable to make quantitative comparisons with
the literature. Furthermore, our results
were supplemented by mean field arguments for the dependence
of $\eta_c(p)$ in the limits $p \rightarrow 1$ and 
$p \rightarrow 0$. 

An additional interesting feature of the model is the
emergence of nontrivial spatial correlations for $p<1$.
We have examined these correlations by an approximate analytic
calculation of the pair distribution function
of a network of objects, and have numerically examined this
function for the case of fibers. Moreover, we have
studied the two-point mass density fluctuation correlation
function of the model, and shown that in some cases
it can be well approximated by a power law type of
form, with a nontrivial exponent. This kind of power
law behavior is enhanced by increasing anisotropy
of the deposited objects. 

Finally,
the results for the correlation function from the model
are particularly interesting when
contrasted with those obtained for mass density correlations in
real paper sheets, where power-law type of
correlations arise in
some cases. Further work
on related, more complicated deposition models for
fibers is in progress.

Acknowledgements: This work has in part been supported by
the Academy of Finland through the MATRA program.
M.J.A. has also been supported by DOE Grant No.
DE-FG02-090-ER45418.

\appendix
\section{The pair distribution function}

In order to derive an expression for $\Omega(r)$ 
we need to know how to perform the ensemble average in
Eq. (\ref{tr}). 
To derive an explicit form for
$G_{\mbox\scriptsize CM}(\vec x_0,\vec x)$ 
we start with a general expression for the
probability density $P[\rho]$ of a given
fiber configuration, where  $\rho$ is the mass
density:

\beq
P[\rho] = \int_A \prod_{k=1}^N d^2\!x_k B(\{\vec{x}_k, \theta_k \}
)\,\delta [\rho(\vec{x}) - \sum_{s=1}^N
\mu(\vec{x},\vec{x}_s,\theta_s)]\,\, ,
\eeq
\noid
where $\mu$ has been defined in Eq. (\ref{massdef})
The 
probability distribution for the centers of masses is readily obtained by
replacing $\mu$ by a delta function.

To calculate
correlation functions the distribution function
of the centers of masses
$B$ has to be provided. In what follows, we will for
simplicity drop the angular dependence from it.
In case we assume that
$p$ is so small that during 
deposition single clusters
emerge and start growing 
independently of each other, the distribution
will factorize as

\beq
B(\{\vec{x}_k \}) = \prod_{m=1}^{n_c} 
B_{\mbox\scriptsize cl}(\{\vec x_s\}_m)\,\, ,
\eeq
\noid
where $\{\vec x_s\}_m$ denotes the set of the center of
mass coordinates of the 
particles belonging to the $m^{\rm th}$ 
cluster, and $n_c$ is the number of the clusters. 
The assumption made here is that each particle must
be uniquely assigned to one of the connected clusters in the network.
The distribution of the particles inside
each cluster $B_{\mbox\scriptsize cl}$ is taken to be a
completely symmetric function of all of its arguments.
In Eq. (\ref{tr})
we will separate the different terms in the correlation function into two
groups. First, there are those terms that come from correlations between
particles belonging to the same cluster (corresponding particle labels are
$i$ and $j$). The remaining
contribution is collected from correlations between particles sitting in
different clusters (labels $I$ and $J$):

\begin{eqnarray}
\lefteqn{}
G_{\mbox\scriptsize CM}(\vec{x}_0, \vec{x}) =  \sum_{i,j}
\int_A \prod_{k=1}^{N} d^2\!x_k B(\{\vec{x}_k \})
\,\delta(\vec{x}_0 - \vec{x}_i)\,\delta(\vec{x}_0
+ \vec{x} - \vec{x}_j) \\
+ \sum_{I,J} \int_A \prod_{k=1}^{N} d^2\!x_k B(\{\vec{x}_k \})
\,\delta(\vec{x}_0 - \vec{x}_I)\,\delta(\vec{x}_0
+ \vec{x} - \vec{x}_J)
\end{eqnarray}

Due to the indistinguishability of the particles the first sum on the
RHS of the previous equation can be replaced with a factor
$\tilde{\omega}_1 \equiv n_c N_c(N_c - 1)/2$. Similarily, the second sum
gives rise to a factor
$\tilde{\omega}_2 \equiv N(N - 1)/2 - n_c N_c(N_c - 1)/2$. Hence,

\begin{eqnarray}
G_{\mbox\scriptsize CM}(\vec{x}_0, \vec{x}) & = & \tilde{\omega}_1 \int_A
\prod_{p} d^2\!x_p B_{\mbox\scriptsize cl} (\{\vec x_p
\})\ \delta(\vec{x}_0 - \vec{x}_i)\,\delta(\vec{x}_0 + \vec{x} -
\vec{x}_j) \\
 & + & \tilde{\omega}_2 \int_A
\prod_p d^2\!x_p B_{\mbox\scriptsize cl} (\{ \vec x_p
\})\ \delta(\vec{x}_0 - \vec{x}_I)
\int_A
\prod_q d^2\!x_q B_{\mbox\scriptsize cl} (\{ \vec x_q
\})\ \delta(\vec{x}_0 + \vec{x} - \vec{x}_J)\\
 & \equiv & \tilde{\omega}_1 B_{\mbox\scriptsize cl}^{(1)}
(\vec{x}_0, \vec{x}_0 + \vec{x})
+ \tilde{\omega}_2 B_{\mbox\scriptsize cl}^{(2)}(\vec{x}_0)B_{
\mbox\scriptsize cl}^{(2)} (\vec{x}_0 + \vec{x})\,\, .
\end{eqnarray}
\noid
In the equation above, particle coordinates $\{x_p\}$ belong to
one of the $n_c$ identical clusters ($\vec x_i, \vec x_j 
\in \{x_p\}$), and $x_q$ belongs
to another cluster ($\vec x_I \in \{x_p\}$ and $ 
\vec x_J \in \{x_q\}$).

The quantity $B_{
\mbox\scriptsize cl}(\{ \vec x_p \})\prod_p d^2\!x_p$ tells the
probability to find the centers of the masses
of the particles within one cluster at
positions $\vec{x}_p$. 
Relative to the center of mass $\vec{x}_{cl}$ of the
cluster these centers of masses cannot be scattered around
the whole system at random positions 
for the cluster is assumed to form a connected structure. From single
cluster growth theory we know that the mass
density $\rho$ is a cone-like function
(given by Eq. (\ref{density})) with radius $R$. 
We assume that the number density
$\eta_{\mbox{\scriptsize CM}}$ of the centers of masses of particles is
proportional to the mass density $\rho$ in the cluster.
The first term 
$B_{\mbox\scriptsize cl}^{(1)} (\vec{x}_0, \vec{x}_0 + \vec{x})$ is
obtained by integrating out the 
positions of all the other centers of masses of particles
but two, which are fixed at the positions
$\vec{x}_0$ and $\vec{x}_0 + \vec{x}$. A nonzero
contribution from the integrations will emerge only from those regions
where the two fixed points are within a distance $R$ 
from the center of mass
$\vec{x}_{cl}$ of the cluster, 
i.e. when they belong to the cluster. Furthermore,
we assume that the number of particles in the cluster is large enough so that
$\vec{x}_{cl}$ is independent of the two points
$\vec{x}_0$ and $\vec{x}_0 + \vec{x}$.
Thus, we can write

\begin{eqnarray}
B_{\mbox\scriptsize cl}^{(1)} (\vec{x}_0, \vec{x}_0 + \vec{x}) & = &
C_1\int_{A} d^2\!x_{cl} \rho(\vec{x}_0 - 
\vec{x}_{cl}) \rho(\vec{x}_0 + \vec{x} -
\vec{x}_{cl})\,\, \label{llq} ;\\
B_{\mbox\scriptsize cl}^{(2)} (\vec{x}_0) & = &
C_2\int_{A} d^2\!x_{cl} \rho(\vec{x}_0 - \vec{x}_{cl}) \label{lt2} \,\, ,
\end{eqnarray}
\noid
where $C_1$ and $C_2$ are normalization constants ($B_{\mbox
\scriptsize cl}$ is a probability density). The
integration domain can be extended over the whole system since $\rho(r) = 0$
when $r > R$. Substitution of
$G_{\mbox\scriptsize CM}(\vec{x}_0, \vec{x})$ into Eq.~(\ref{ty1}) yields

\beq
\Omega(r) dr  =  \omega_1 g(r) dr + \omega_2 \Omega_h(r,L)
dr\,\, ,
\eeq
\noid
where $\omega_1 \equiv \tilde{\omega}_1\frac{2}{N(N-1)}$, $\omega_2 \equiv
\tilde{\omega}_2\frac{2}{N(N-1)}$, and

\begin{eqnarray}
g(r) dr& = & \int_A d^2\!x_0
B^{(1)}_{\mbox\scriptsize cl} (\vec{x}_0, \vec{x}_0 + \vec{x})
\Delta A(\vec{x}_0 + \vec{x}) \label{tk} \\
& = & 2\pi r dr C_1 L^2 \int_{K} d^2\!x_0 \rho(\vec{x}_0 - \vec{x}_{cl})
\rho(\vec{x}_0 + \vec{x} - \vec{x}_{cl});
\label{gex} \\
\Omega_h(r,L) dr & = & C_2^2 \int_A d^2\!x_0 \int_{A} d^2\!x_{cl}
\rho(\vec{x}_0 - \vec{x}_{cl}) \int_{A} d^2\!x_{cl}'
\rho(\vec{x}_0 + \vec{x} - \vec{x}_{cl}') \Delta A(\vec{x}_0 + \vec{x})
\label{tk2} \\
& = & \int_A d^2\!x_0 \Delta A(\vec{x}_0 + \vec{x}).
\label{tk3}
\end{eqnarray}
\noid
where the integration domain $K$ in Eq. (\ref{gex}) is
given by $\vert \vec{x}_0 - \vec{x}_{cl} \vert \le R$,
and $\vert \vec{x}_0 + \vec{x} - \vec{x}_{cl} \vert \le R$,
which is the region where there in a nonzero contribution
to $g(r)$. This means that we can make the transformation
of variables
$\vec{u}=\vec{x}_0 - \vec{x}_{cl}$, 
$2\vec{v} =  \vec{x}_0 + \vec{x}_{cl}$, and thus 
the final result only depends
on $r\equiv \vert \vec{x} \vert$. 

In $d = 2$ $C_1 = (L\pi R^2 a /3)^{-2}$ and $C_2 = (L^2\pi R^2 a /3)^{-1}$
with $a \equiv \rho(0,N)$. The density $\rho$ is given in Eq. (\ref{density}).
Plugging it into Eq. (\ref{gex}) and setting $\Delta A(\vec{x}_0 +
\vec{x}) = 2\pi r dr$, which is true within each cluster, yields
$g(r)$. To get some insight how $g$ behaves we will use one dimensional
density profiles instead of two dimensional ones. However, we do not replace
$\Delta A(\vec{x}_0 + \vec{x})$ with its one dimensional counterpart,
$\Delta A(\vec{x}_0 + \vec{x}) = 2\, dr$ since we want to satisfy the
relation $g(0) = 0$, which is true in $d = 2$. In this case, the
normalization constant $C_1$ is given by
$C_1 = 15/(7 L^2 \pi R^3 a^2)$. This gives as the final
result the Eq. (\ref{gog}).

Finally, we would like to note that in
Eq. (\ref{tk3}) $\Delta A(\vec{x}_0 + \vec{x})$ imposes a
cutoff at the edges of the system. This area element can
be written in the form

\beq
\int_A d^2\!x_0 \Delta A(\vec{x}_0 + \vec{x}) = r dr \int_A d^2\!x_0
\beta(\vec x_0, \vec x),
\eeq
\noid
where $\beta(\vec x_0, \vec x)$ tells the maximum angle swept by a vector 
of length $r = |\vec x|$ rotated around a fixed position 
$\vec x_0$, with the constraint that the vector must lie inside 
the system. This can be used in deriving 
the pair distribution function of a uniformly random
system $\Omega_h$, as given by Eq. (\ref{unif}).


\newpage 

FIGURE CAPTIONS

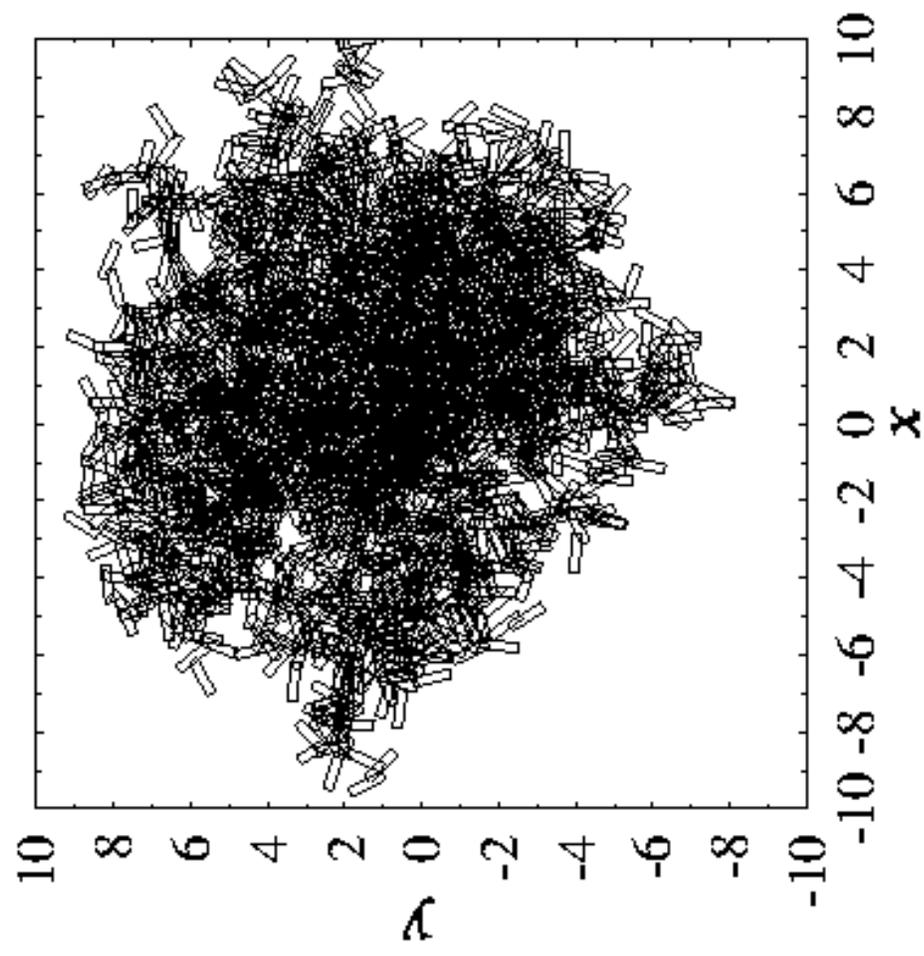
\noindent Fig 1. 
A snapshot of a single growing fiber cluster for $N=2000$ fibers 
of length $\lambda=1$ and width $\omega=1/4$. \\

\noindent Fig 2. Snapshot of a disordered network of $N=2000$ 
deposited needles of length 
$\lambda=1$, for
(a) $p=1.0$, (b) $p=0.1$, and (c) $p=0.001$. \\

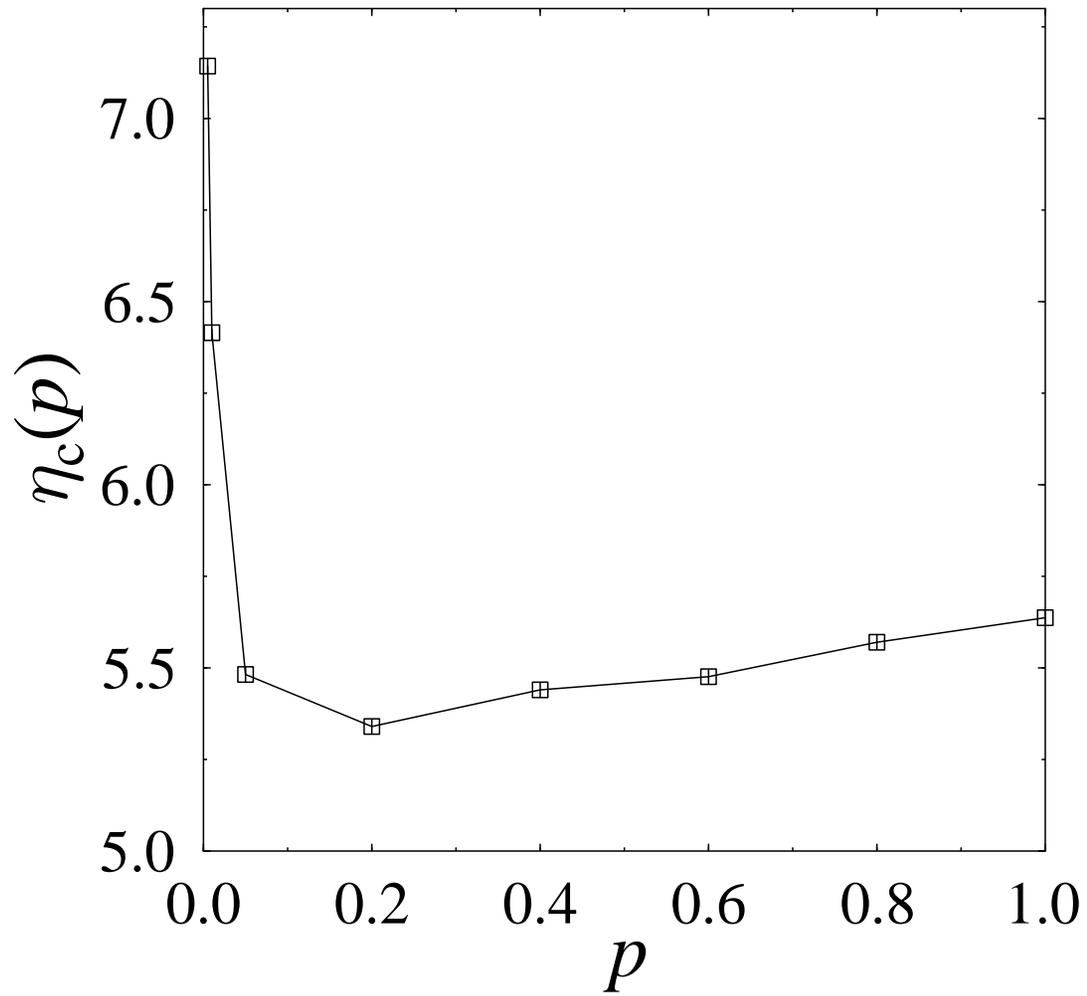
\noindent Fig 3.
The critical percolation threshold $\eta_c(p)$ vs. $p$ for 
a network of needles of length $\lambda=1$, 
extrapolated to $L \rightarrow \infty$. The error
bars are of the size of the points. \\

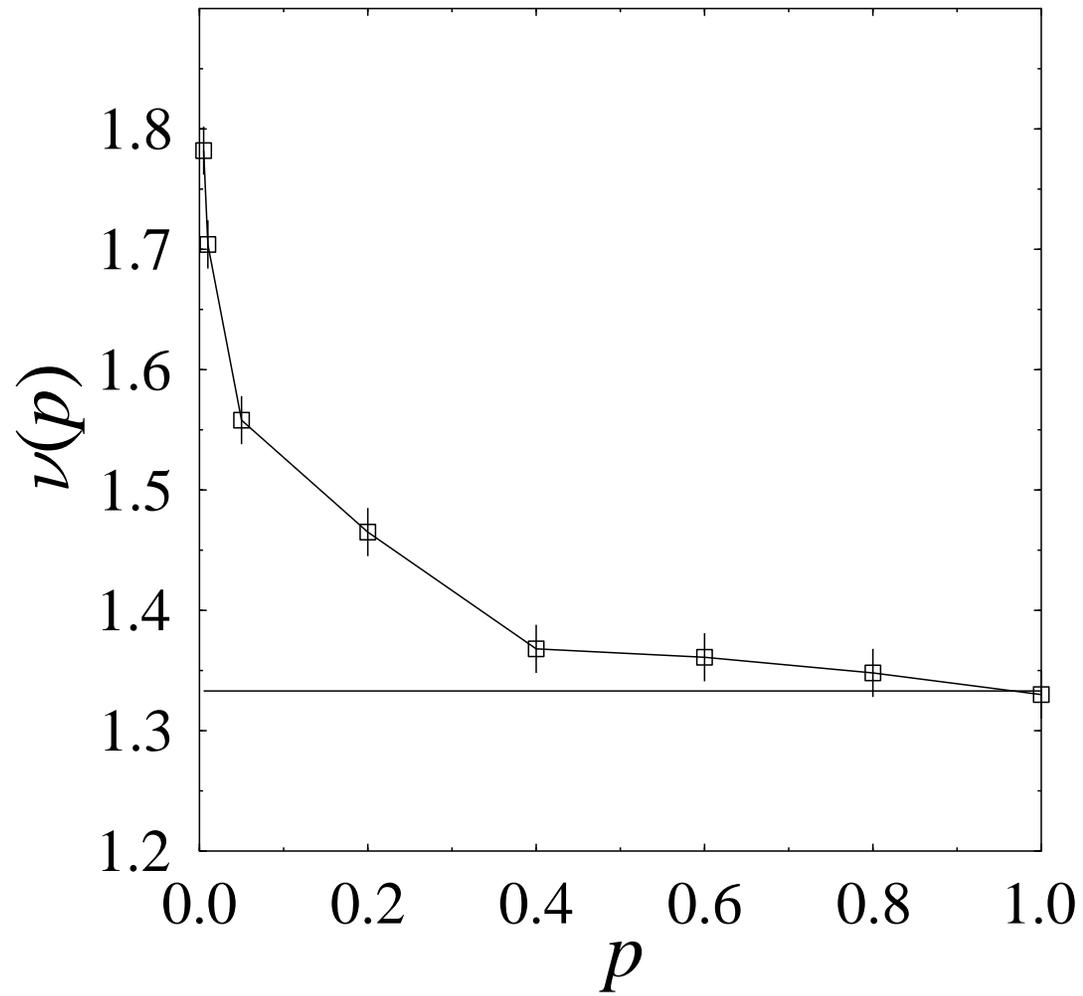
\noindent Fig 4.
Correlation length exponent $\nu(p)$ vs. 
$p$ for needles of length $\lambda=1$. In this and the following
figures for $\nu(p)$, the horizontal line denotes the exact value
$\nu=4/3$. \\

\noindent Fig 5. 
Snapshot of a disordered network of $N=2000$ fibers 
of length $\lambda=1$ and 
width $\omega=1/4$. In (a) $p=1.0$, in (b) $p=0.1$, and in (c) $p=0.001$. \\

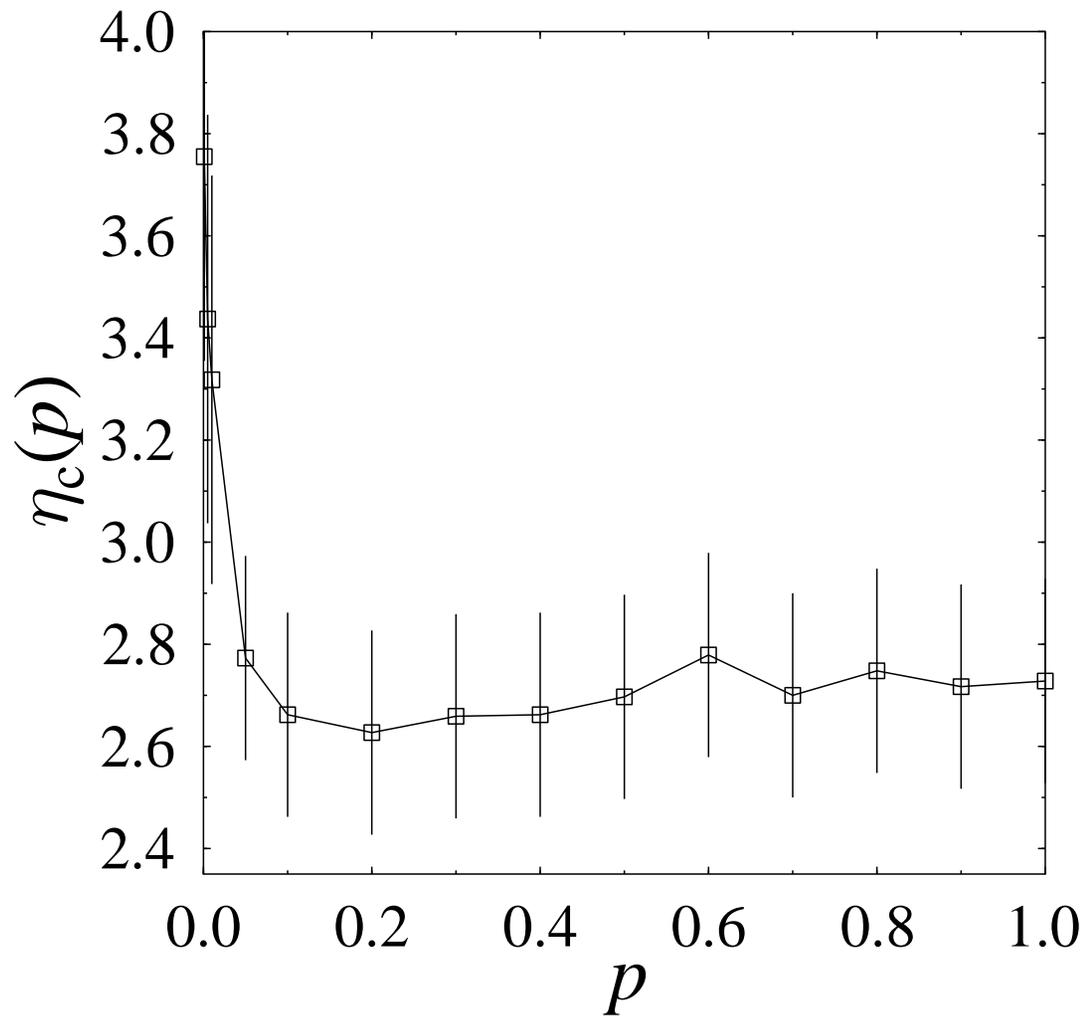
\noindent Fig 6.
The critical percolation threshold $\eta_c(p)$ vs. $p$ for a 
network of fibers of length $\lambda=1$ and 
width $\omega=1/4$, extrapolated 
to $L \rightarrow \infty$. See text for details.\\

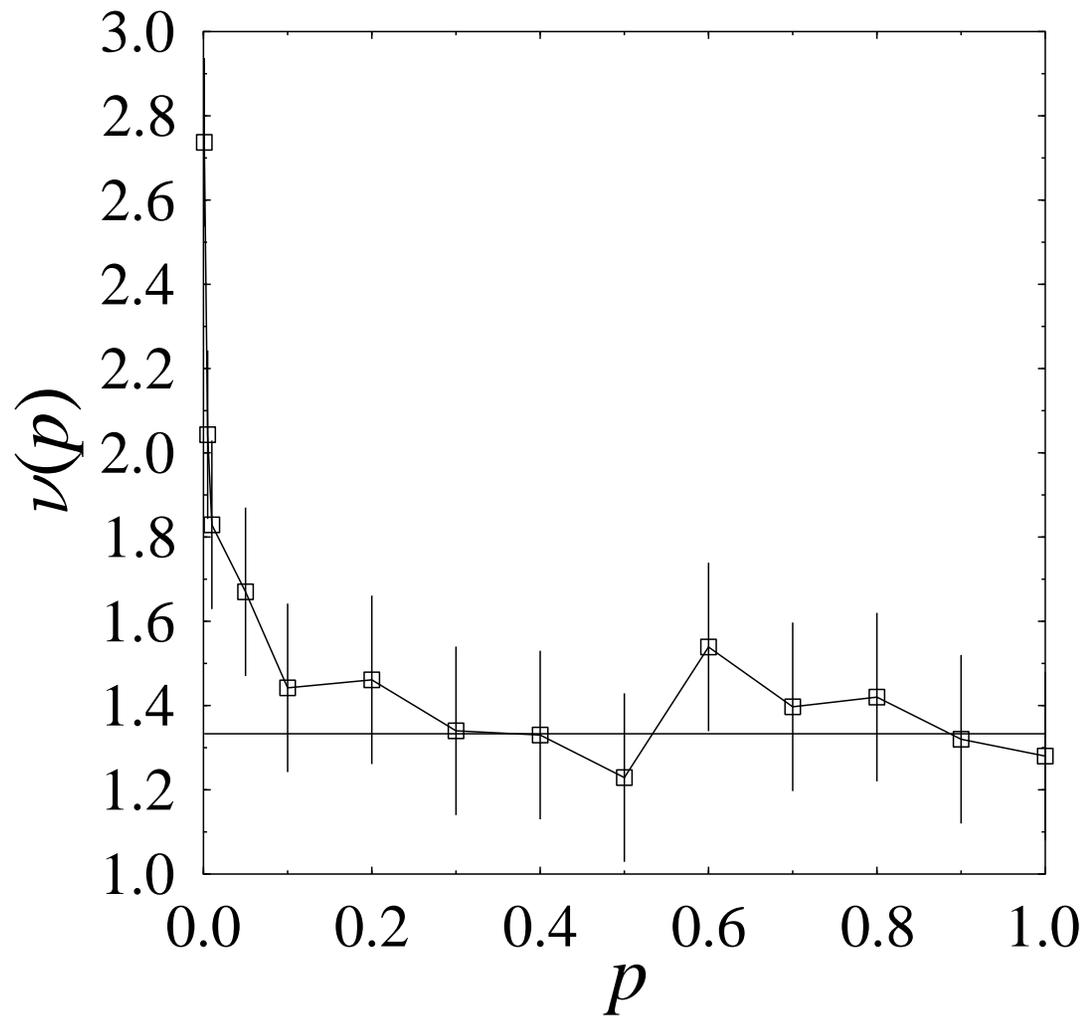
\noindent Fig 7.
Correlation length exponent $\nu(p)$ 
vs. $p$ for fibers of length $\lambda=1$ and 
width $\omega=1/4$.\\
  
\noindent Fig 8. 
Snapshot of a network of $N=1000$ disks of radius $r_d=0.5$ in a system of
size $L=20$.
In (a) $p=1.0$, in (b) $p = 0.1$, and in (c) $p=0.001$. \\

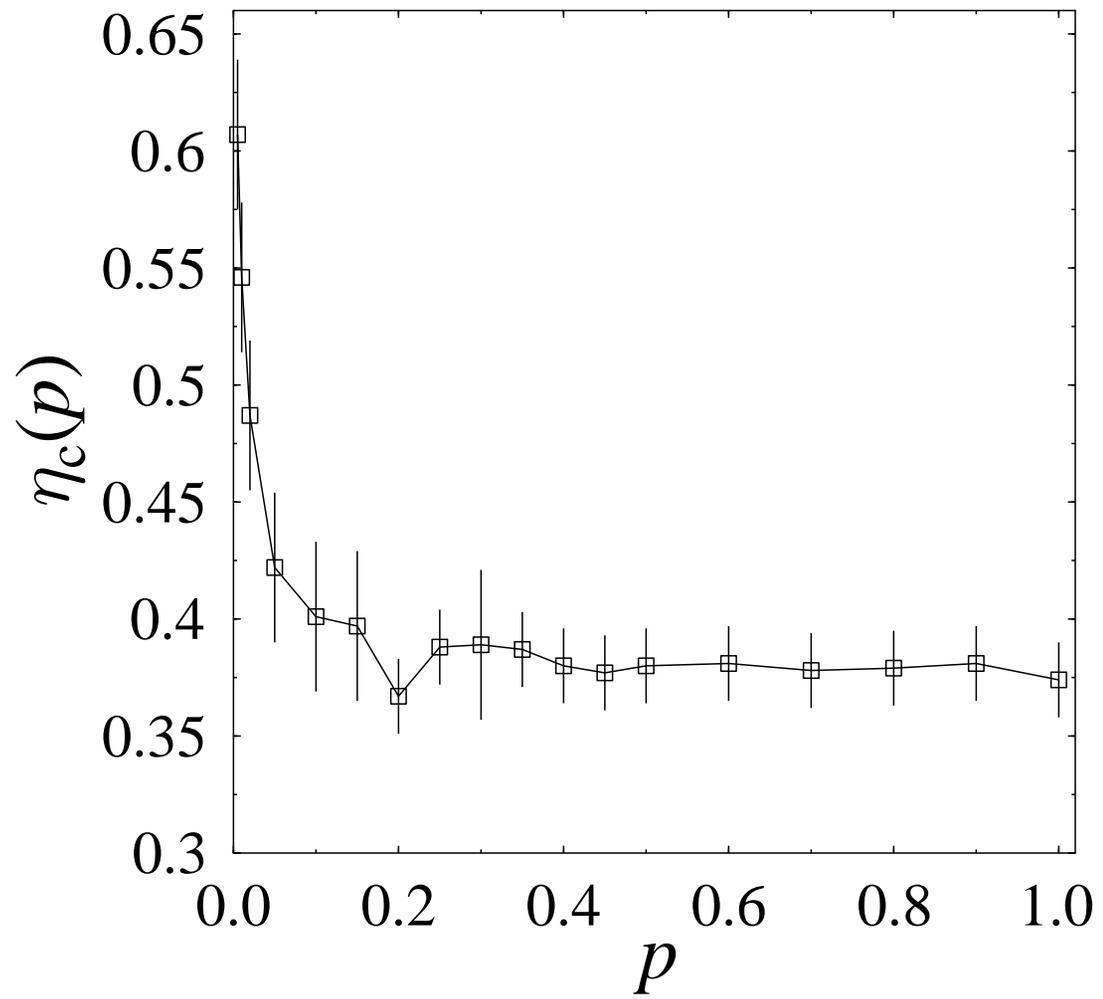
\noindent Fig 9.
The critical percolation threshold $\eta_c(p)$ vs. $p$ for a network 
of disks of radius $r_d=1$, extrapolated to $L \rightarrow \infty$.\\

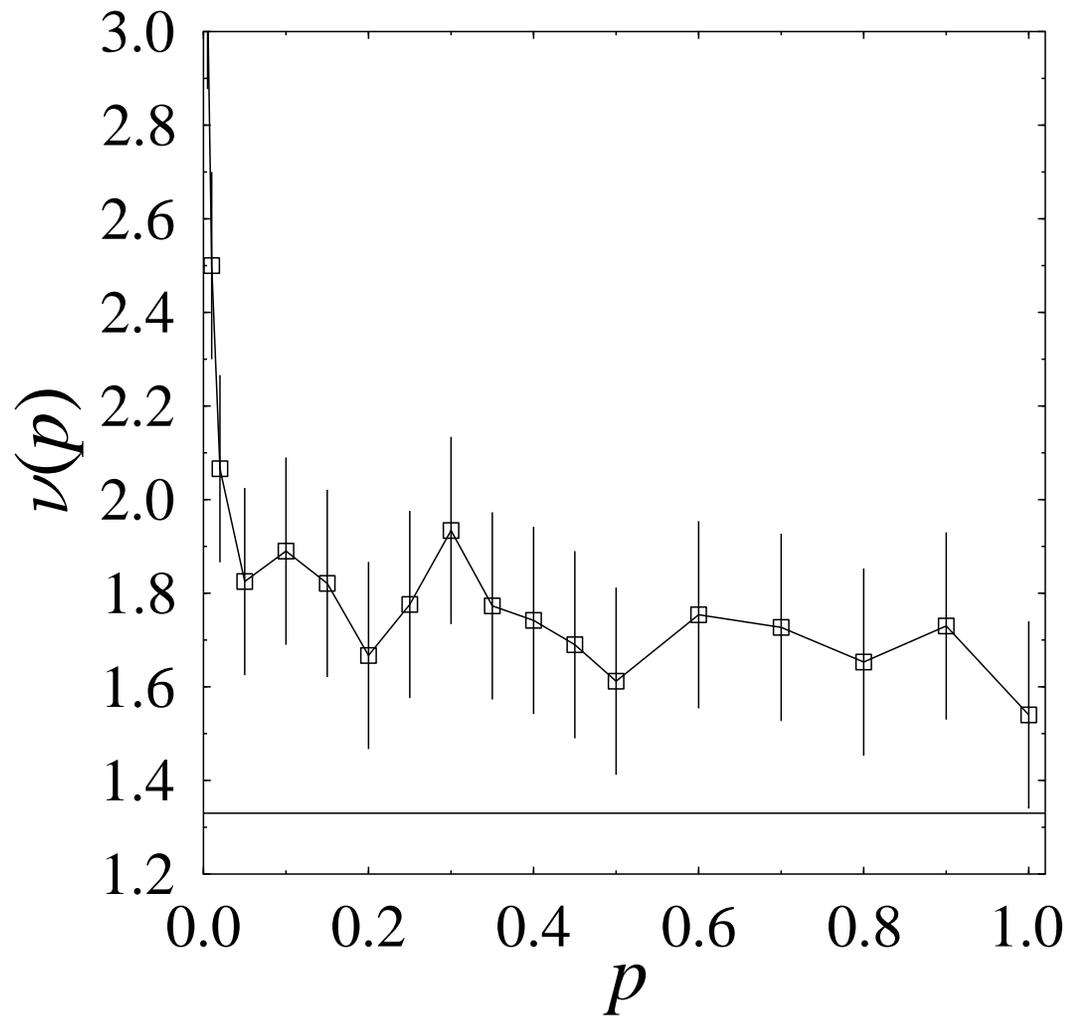
\noindent Fig 10.
Correlation length exponent $\nu(p)$ vs. $p$ 
for disks of radius $r_d=1$. \\

\noindent Fig 11. 
Simulated pair distribution functions for 
fibers of length $\lambda=1$ and 
width $\omega=1/4$ for (a) $p=0.001$ and (b) $p=0.01$.
The different curves correspond to 
$N=15, 25, 75,150, 250,500,750,1000,2000,4000,4500,
5000, 6000$ from top to bottom at $r=0$.\\

\noindent Fig 12.
Comparison between analytic (solid lines) and simulated (dotted lines)  
pair distribution functions for fibers of length $\lambda=1$ and 
width $\omega=1/4$, for $p=0.001$ and $L=20$. 
In (a) $N=250$, (b) $N=500$, and
(c) $N=750$. See text for details in this and the following
figures.\\

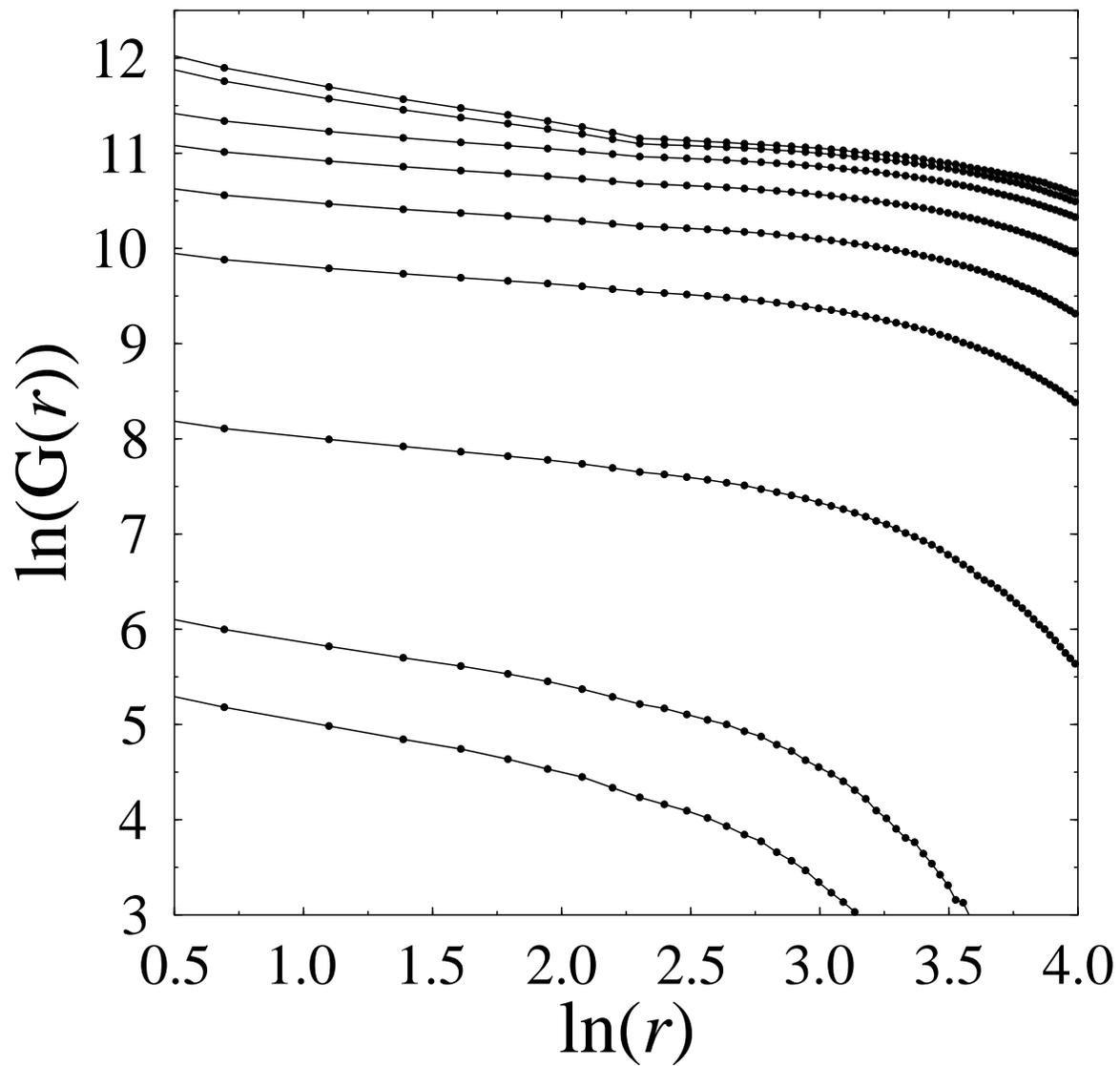
\noindent Fig 13. 
A plot of $\ln(\tilde{G}(r))$ vs. $\ln(r)$ for fibers of length 
$\lambda=20$ and width $\omega=1$, for $p=0.001$. From 
bottom to top, $N$ increases with 
$N=25, 50, 300, 1500, 3000, 5000, 8000, 20000, 25000$.
The system size $L=400$. \\

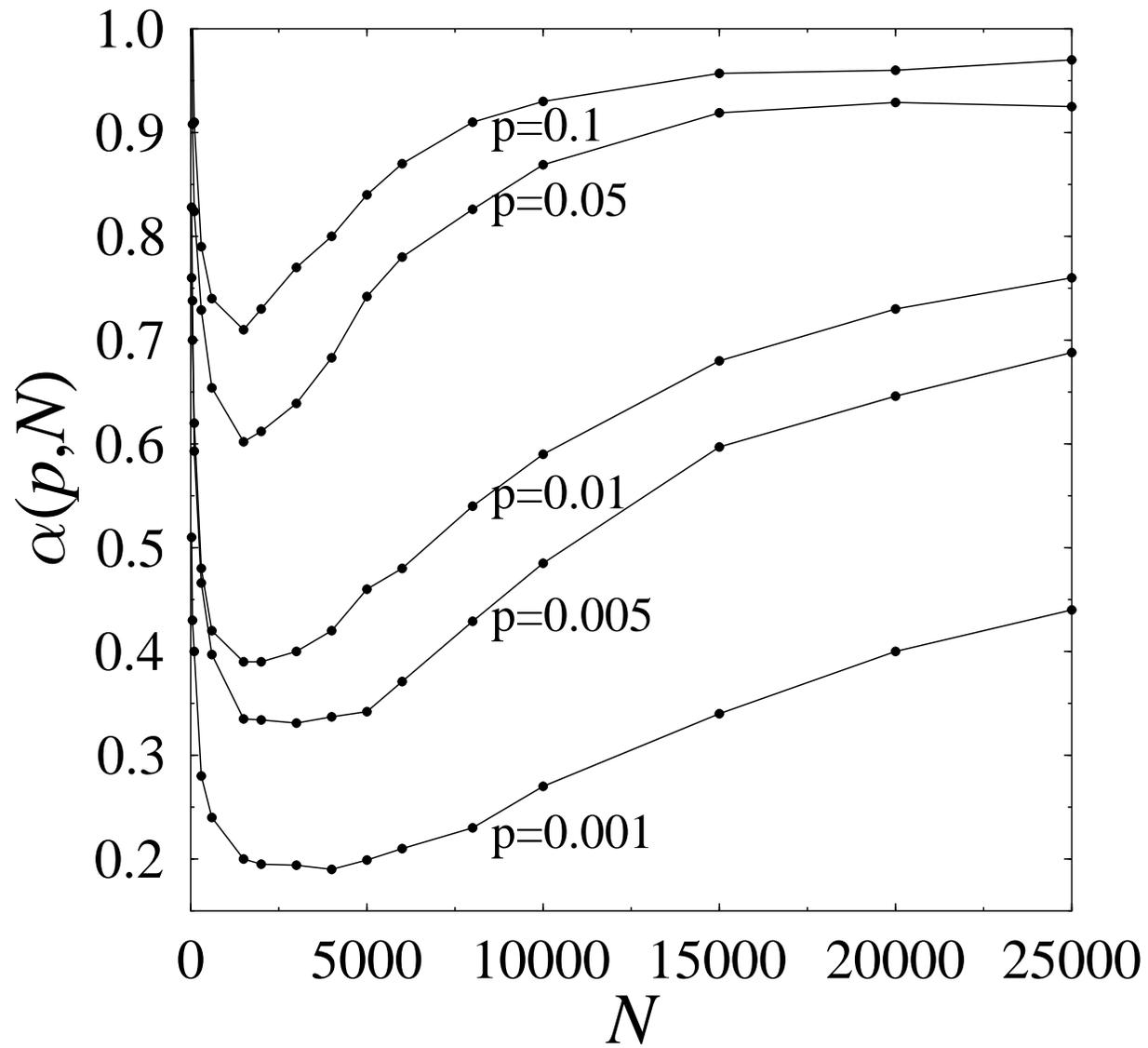
\noindent Fig 14.
A plot of the effective exponent $\alpha(N,p)$
vs. $N$ for various $p$.  
The aspect ratio is $L/\lambda/\omega =400/20/1$.
\\

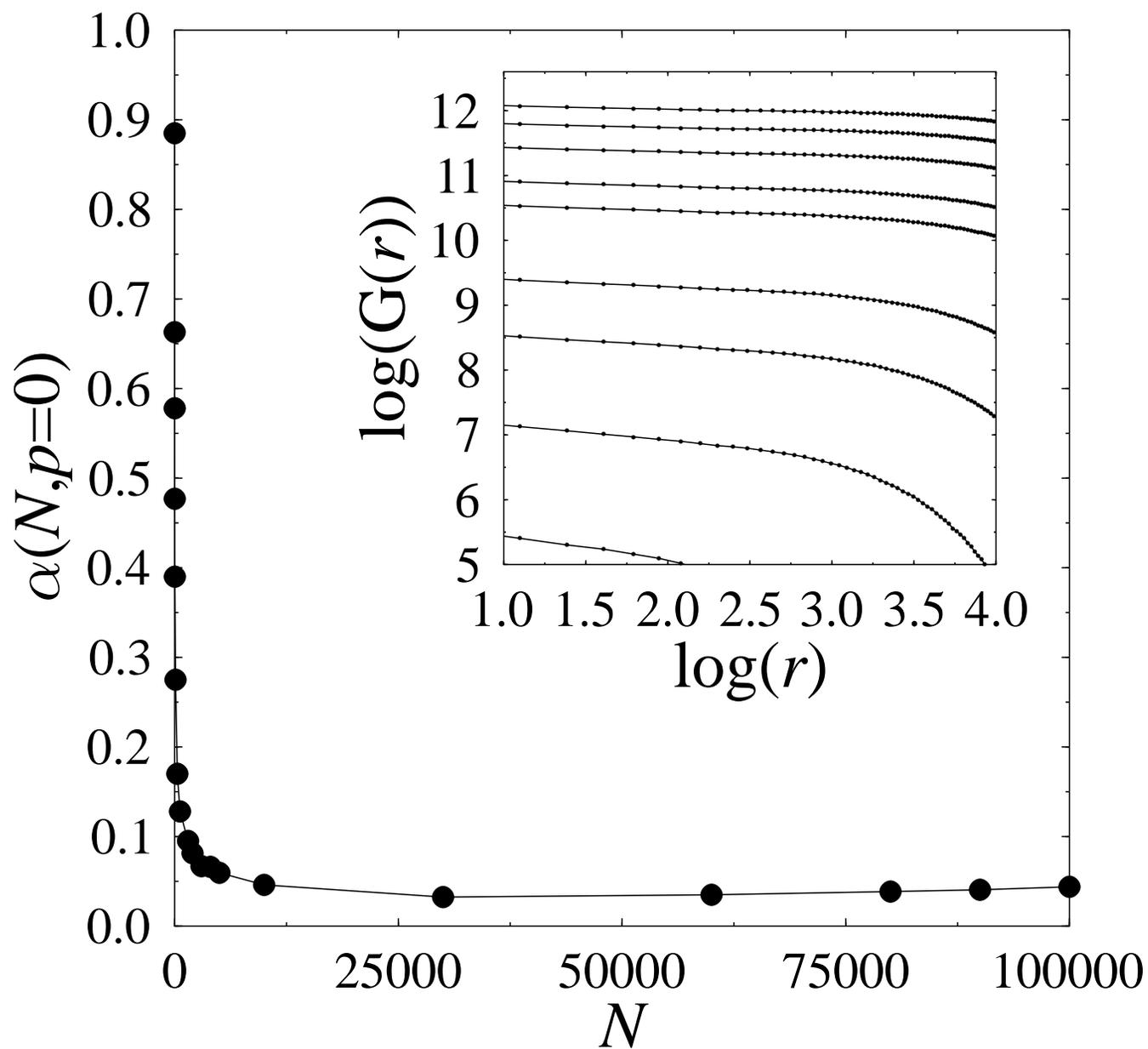
\noindent Fig 15.
A plot of $\alpha(N,p=0)$ vs. $N$ for the special case 
where a single cluster grows with fibers of length 
$\lambda=20$ and width $\omega=1$ (see Fig. 1). Note 
the saturation to a constant $\alpha\approx 0.05$ 
after $N \approx 500$. The inset shows $\ln(\tilde{G}(r))$
vs. $\ln(r)$ for $N=25,100,300,600,1500,2000,3000,4000,
5000$, from bottom to top. \\


\begin{thebibliography}{99}
\bibitem{Timo}
This model was suggested to us by Jussi Timonen and
originally introduced in Ref. \cite{Ast89}.
\bibitem{box}
We have checked the density profiles explicitly as a function
of $p$ and find that near $p=1$, $L'$ need not be larger than
the largest object dimension. However, for the smallest values
of $p$ studied here we find that $L'$ must be about five times
larger.
\bibitem{boundary}
According to Ref. \cite{Rob84}, if the critical number
density is taken to be
the average of the values obtained for the
one and two-sided cases, the finite-size correction
should approximately cancel out. We checked this for needles, but
found it not to be true.
\bibitem{excluded}
The excluded volume $\langle A \rangle$ can be evaluated from
Eq. (19) of Ref. \cite{Bal84} by using the present fiber dimensions
and setting $\theta_{\mu}=\pi/2$.
\bibitem{onesided}
In the particular case of disks, the spanning
probabilities were obtained by fixing one direction
and searching for the percolating cluster in that direction
only.
\bibitem{clusters}
To verify the argument that for $p$ slightly less than unity,
the dominant effect is the removal of single needles
from the network, 
we also calculated numerically the cluster size
distributions. Since also doublets, triplets etc. are
removed, the analytic theory underestimates the
slope of the curve for $p<1$.
\bibitem{fnote}
As a consistency check, we have explicitly verified that
$\tilde{G}(r)$ as calculated from the analytic theory 
agrees with simulations of the lattice model for
$p=0.01$ and over a
range of $N$ for the choice of $L/\lambda/\omega=400/20/1$.
We solved our lattice model with periodic boundary conditions.
For the system sizes used we found that the effect of the boundaries
is not detectable.
\end{thebibliography}
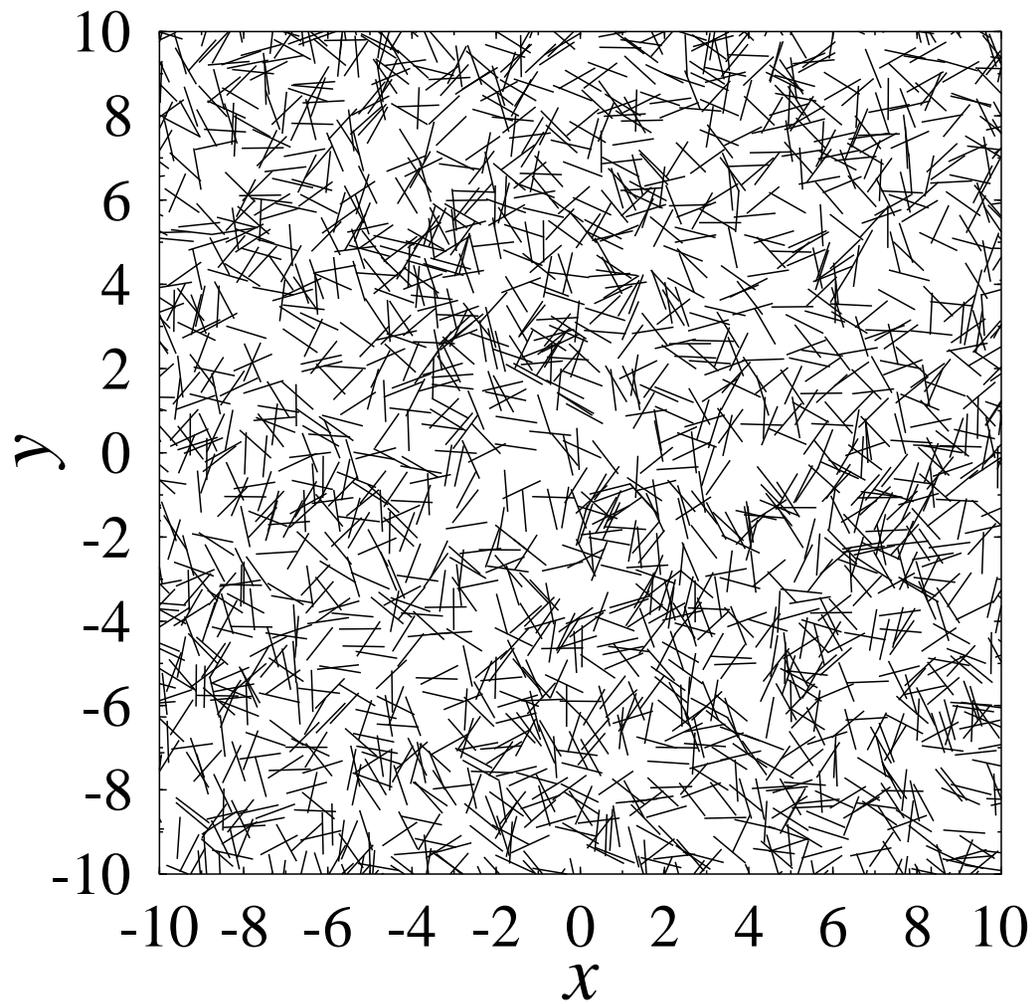
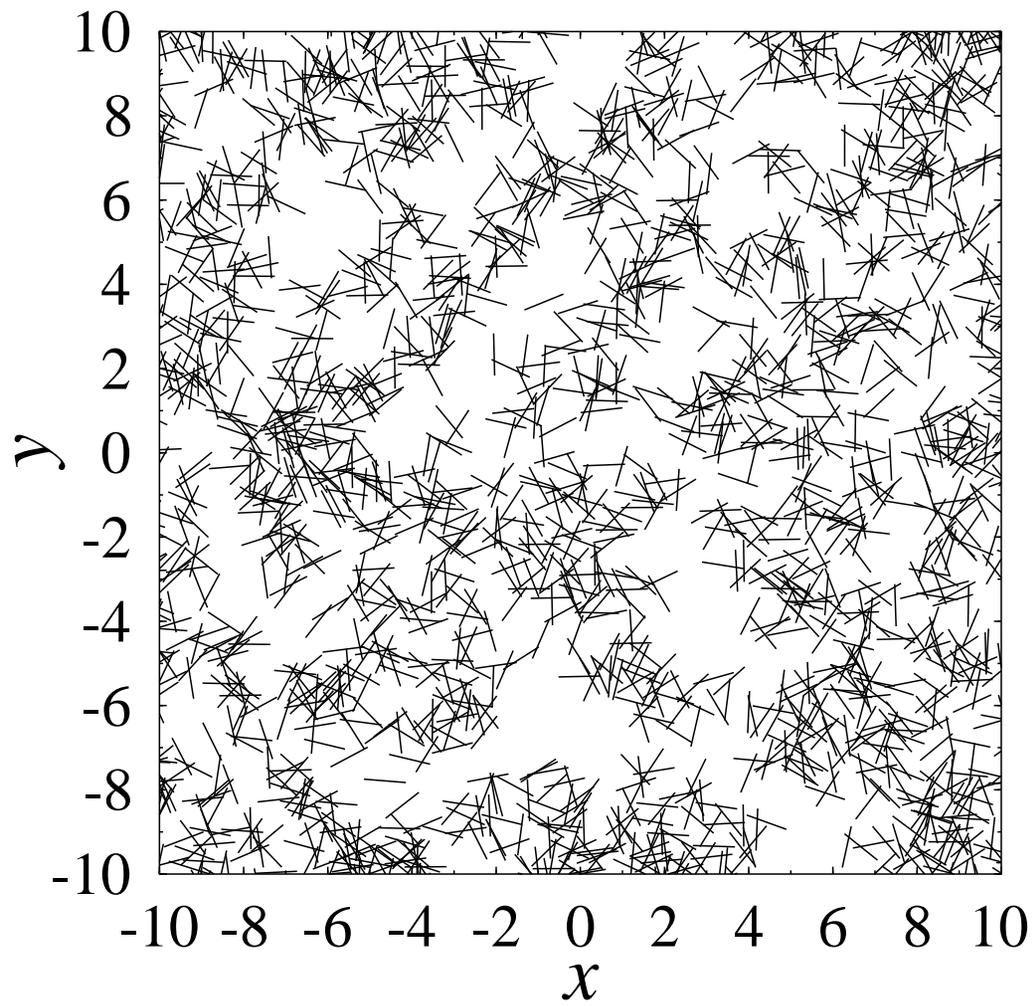
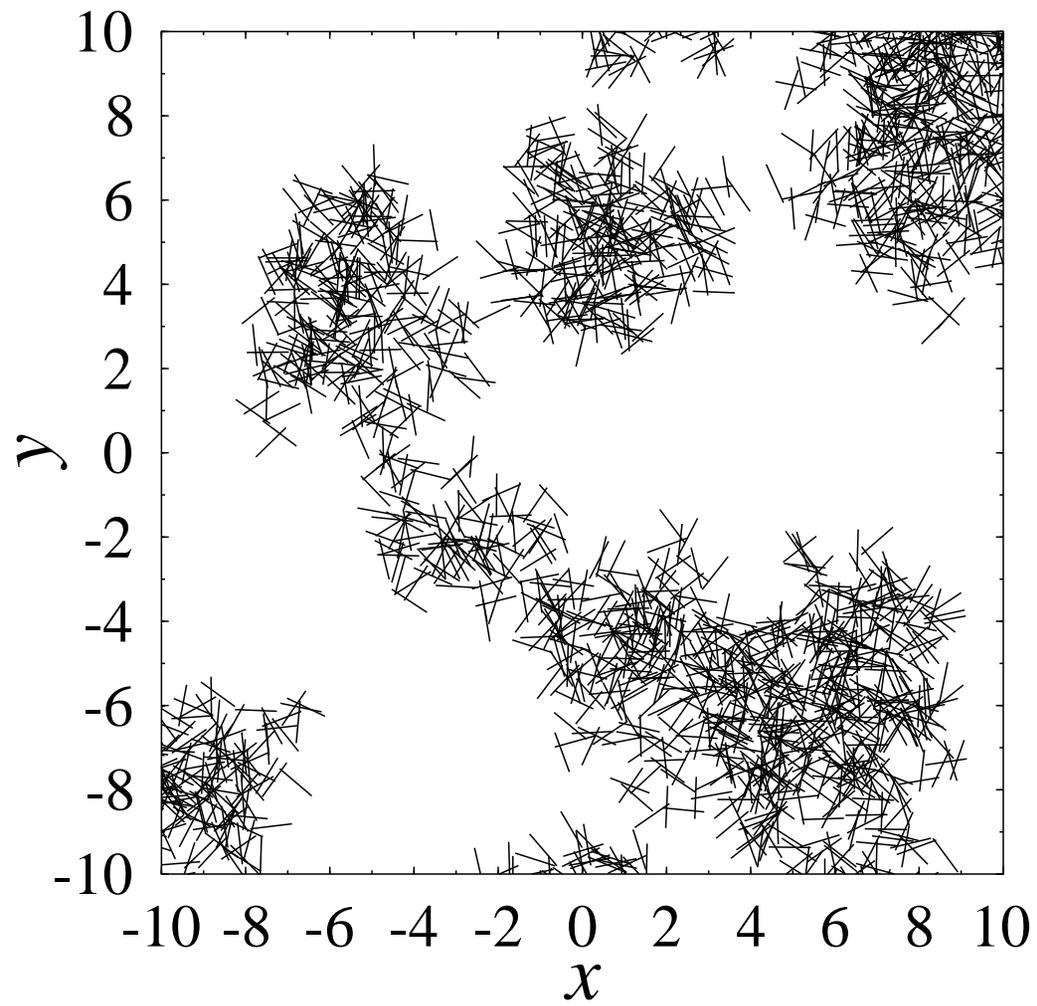
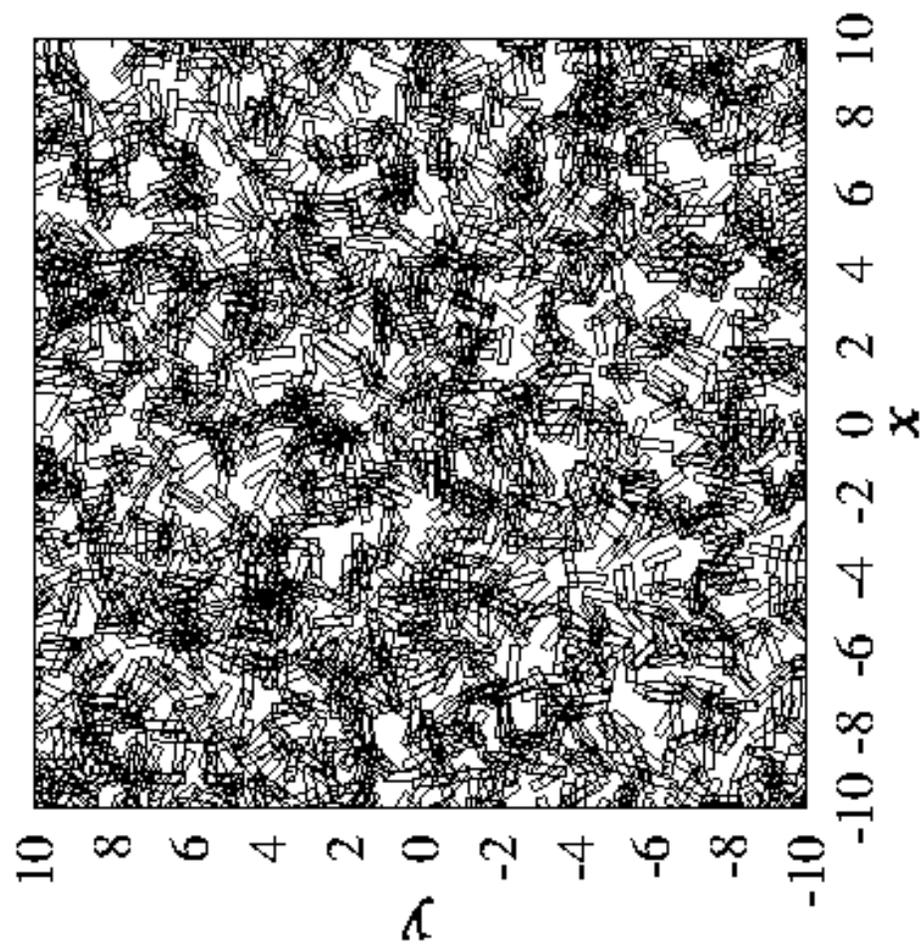
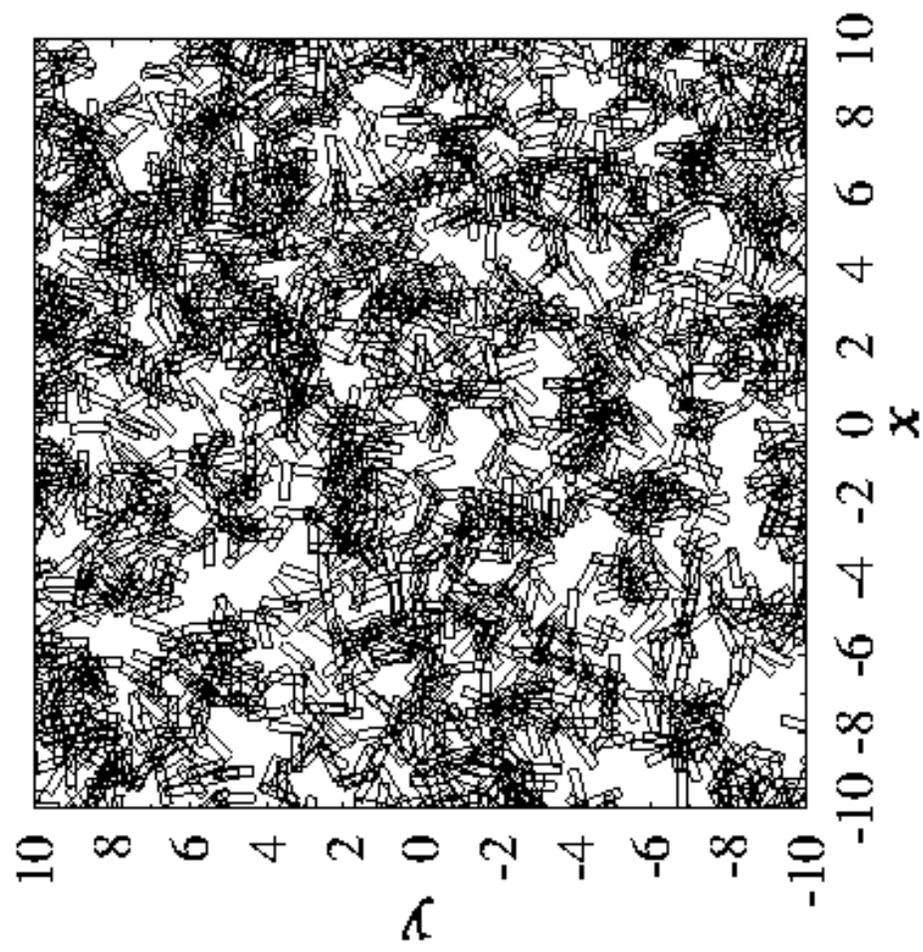
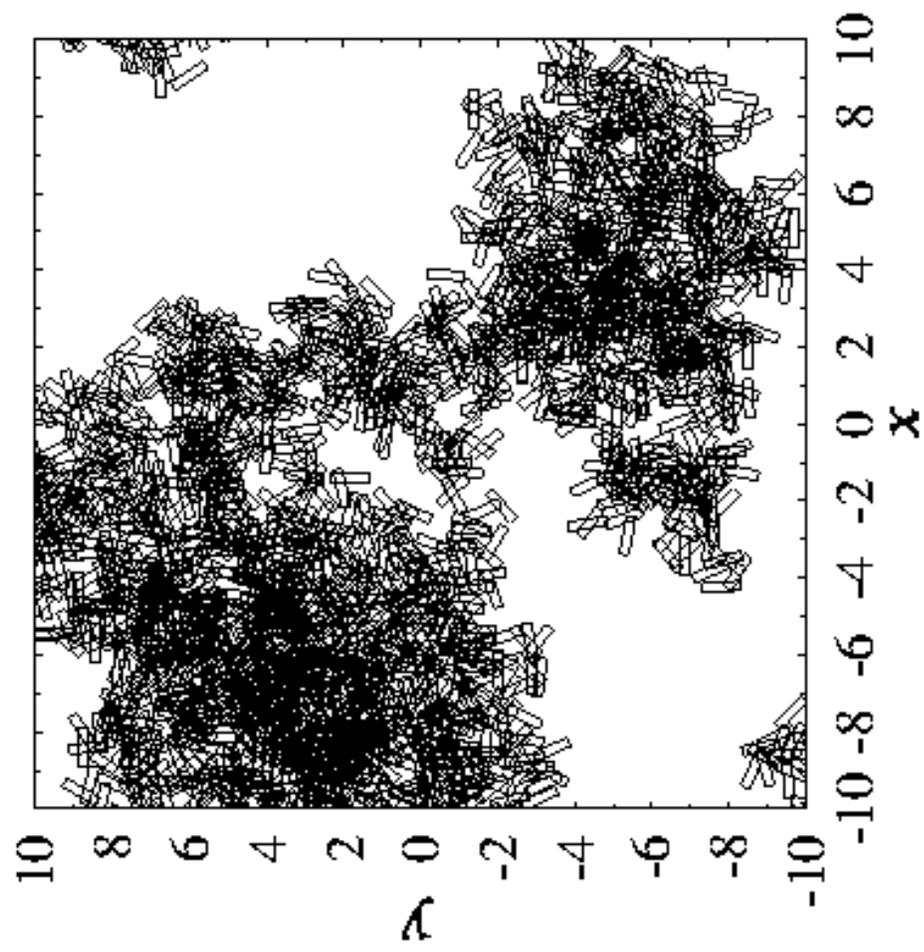
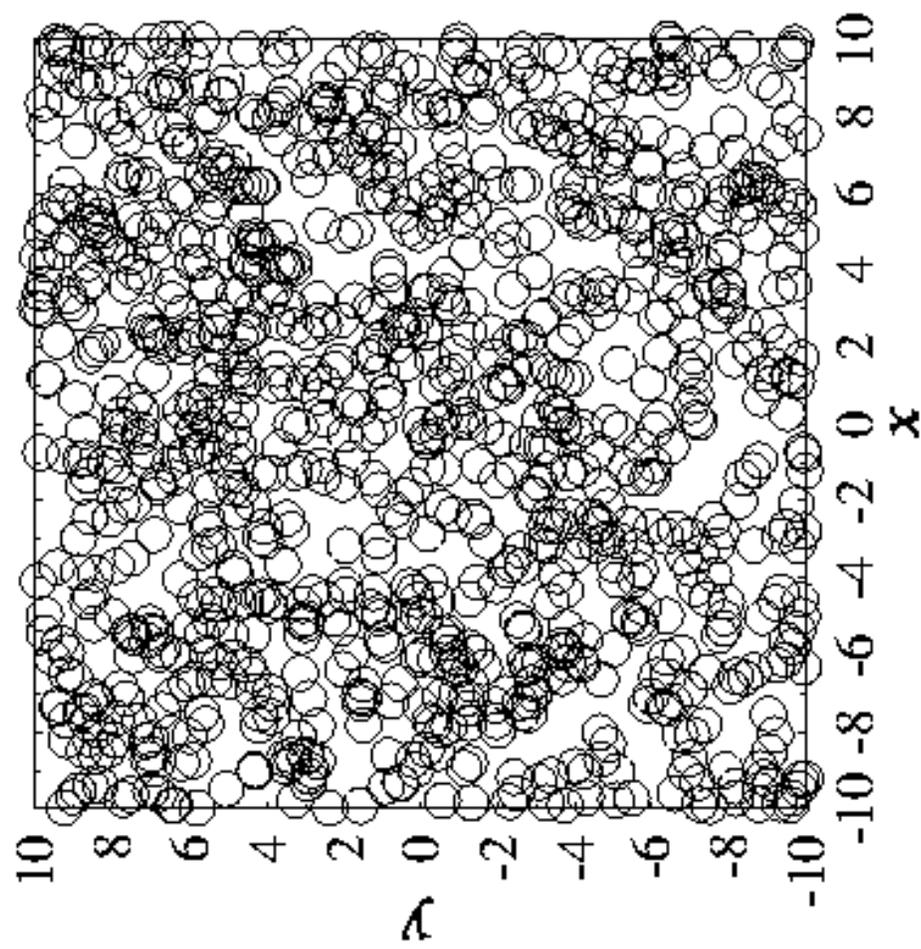
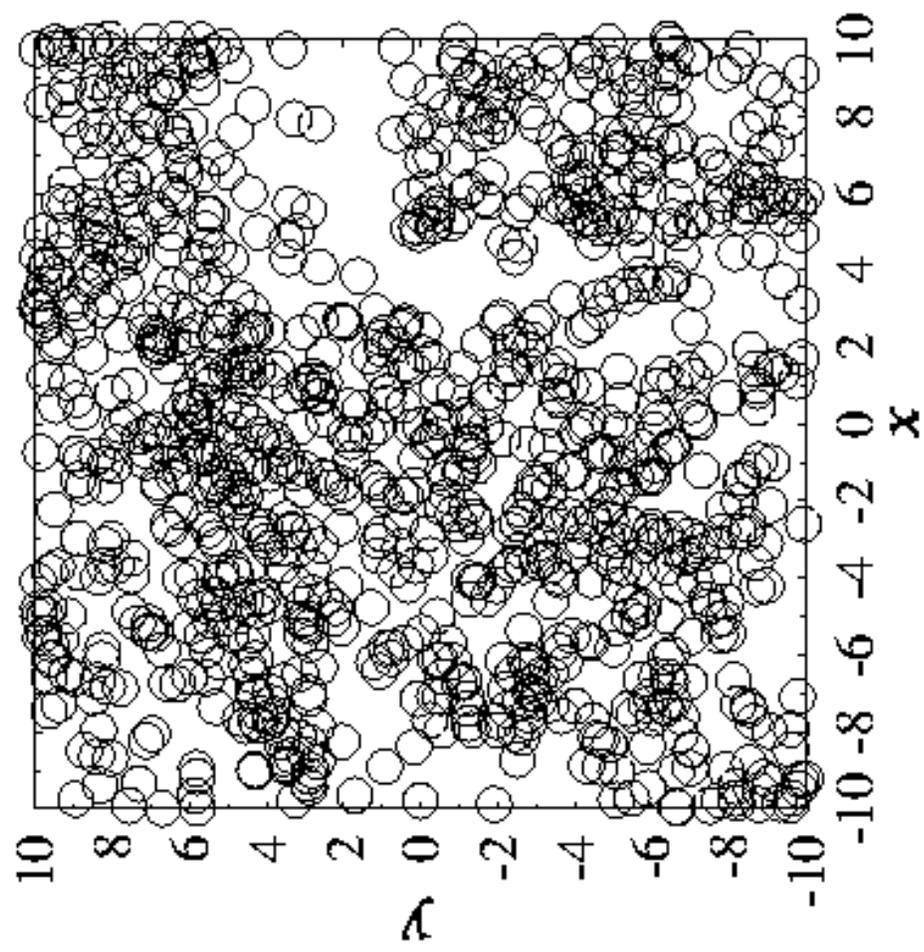
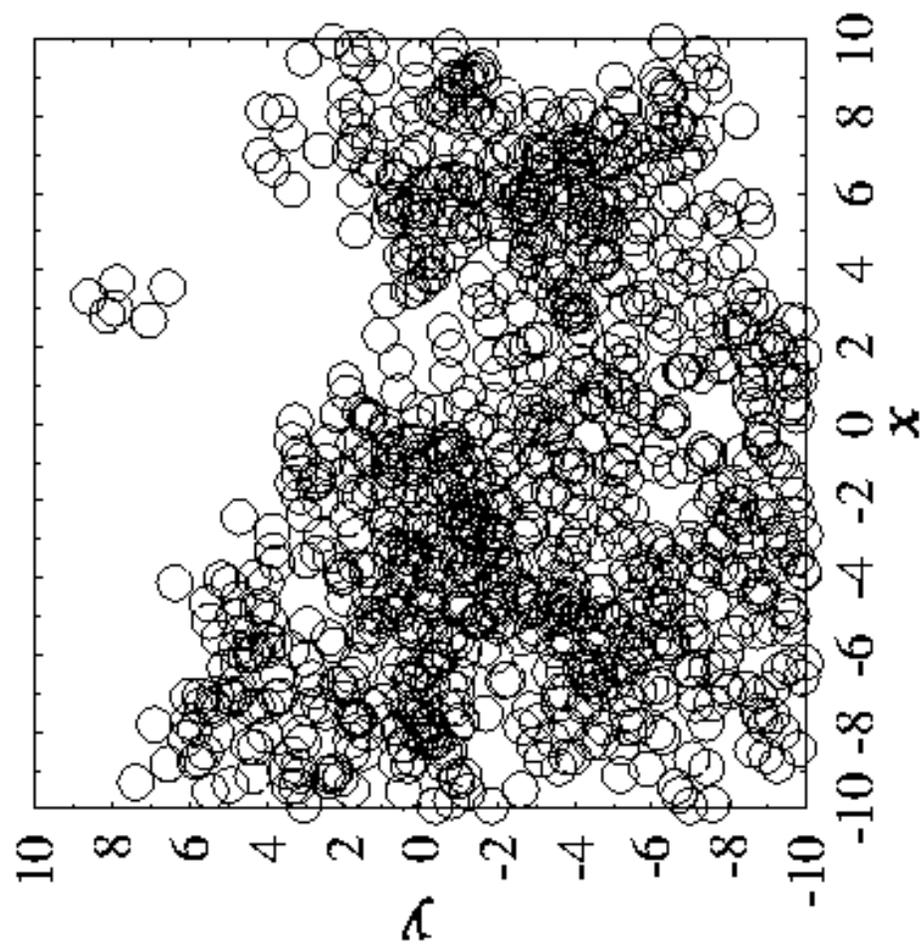
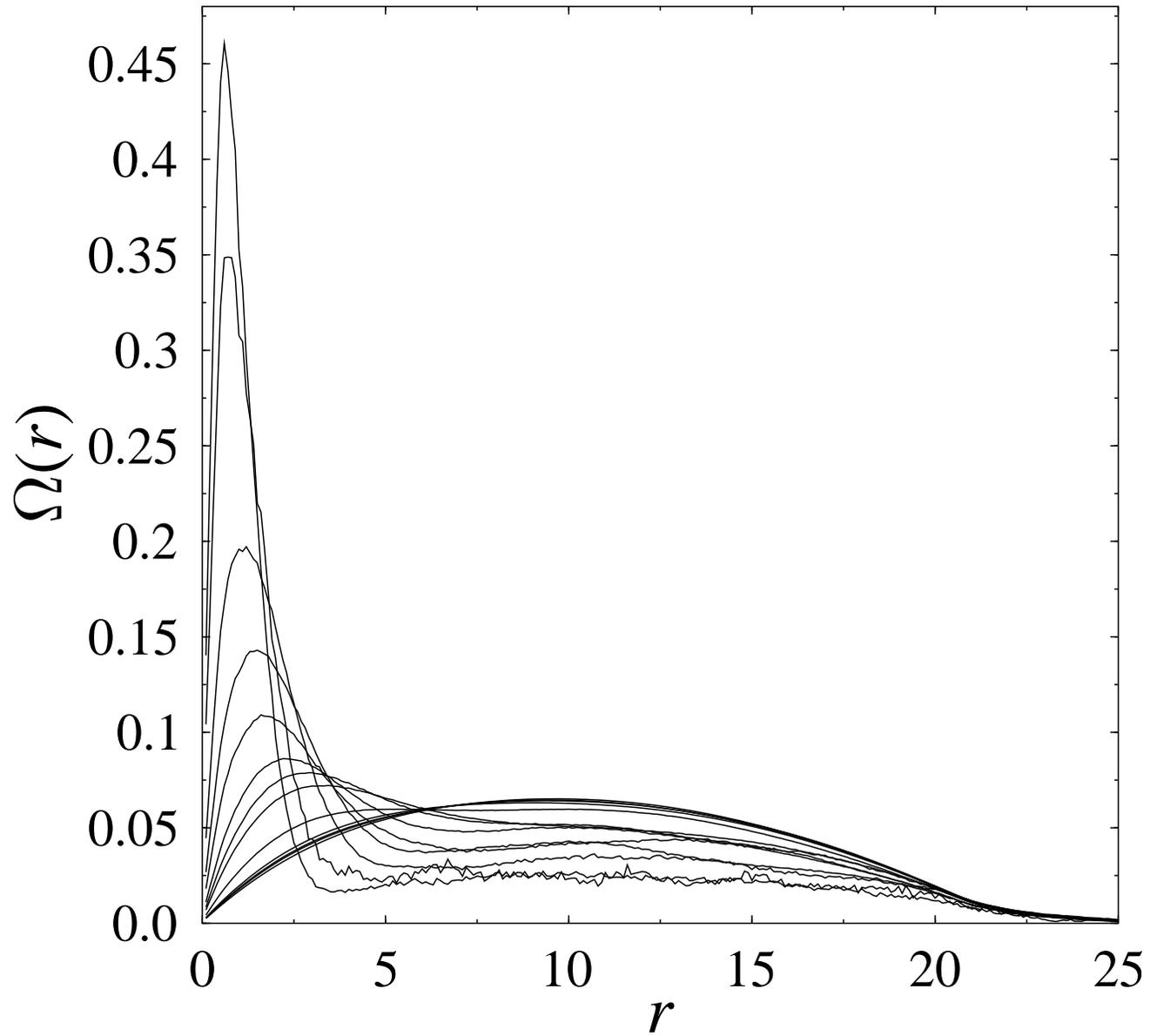
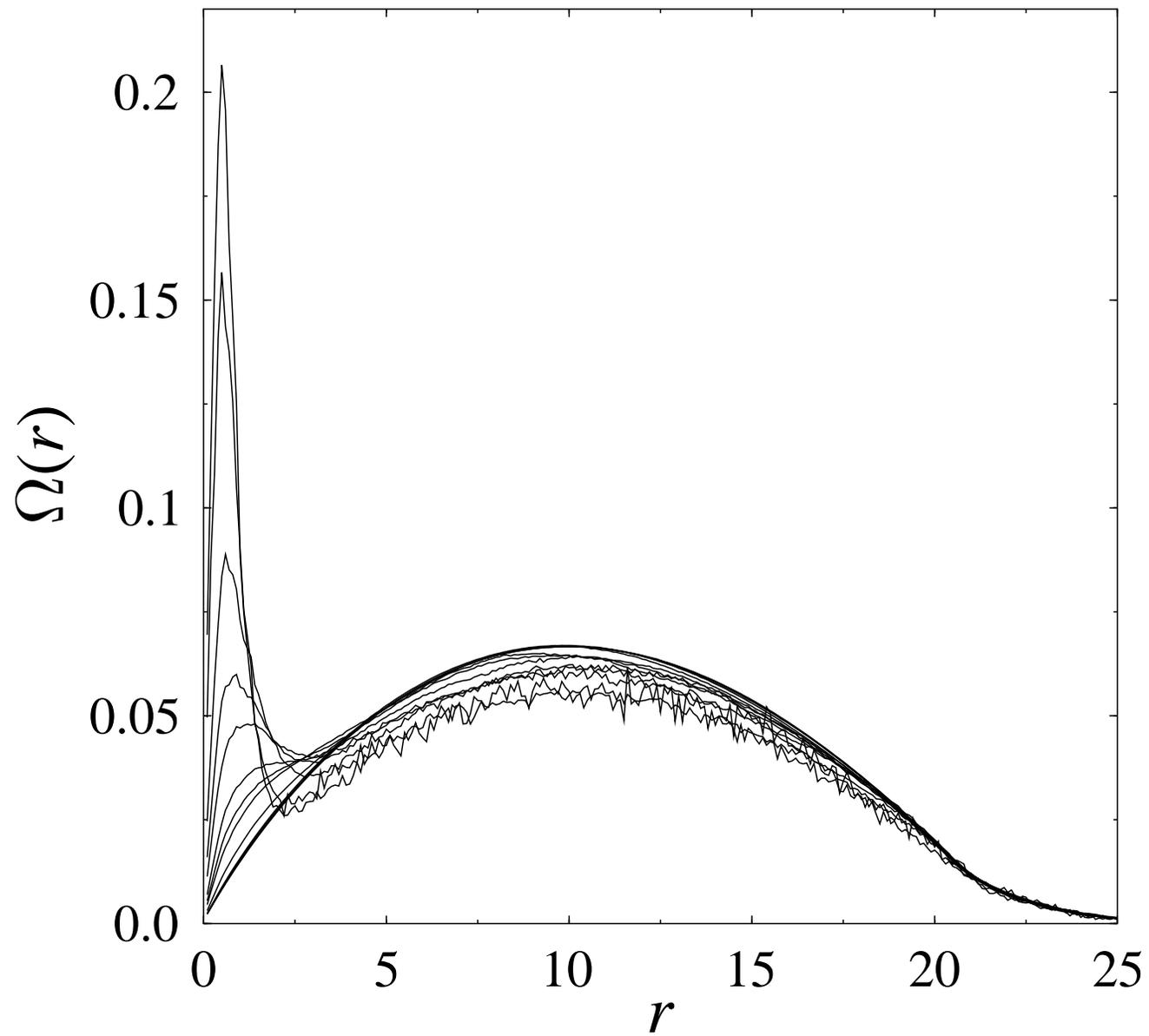
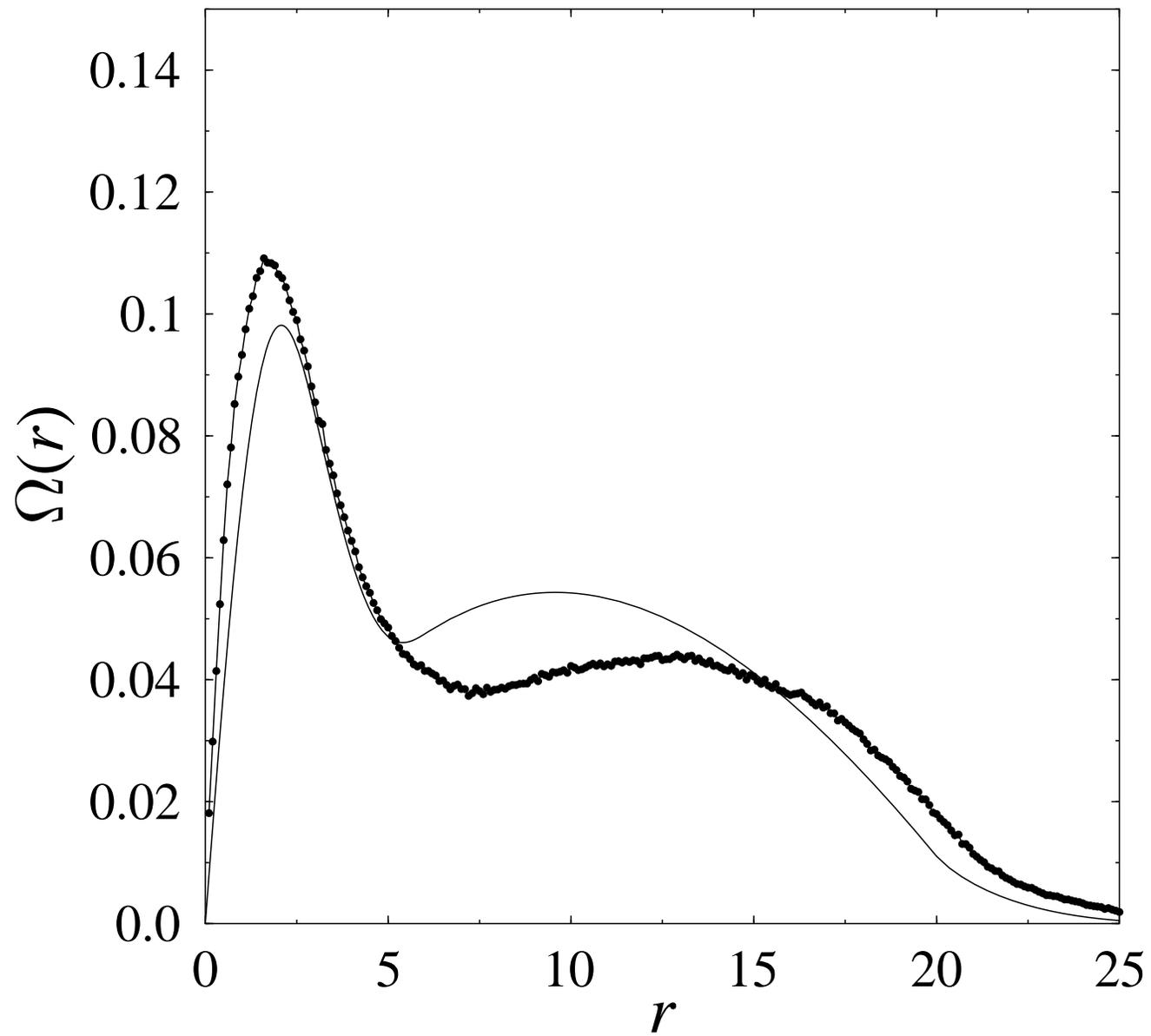
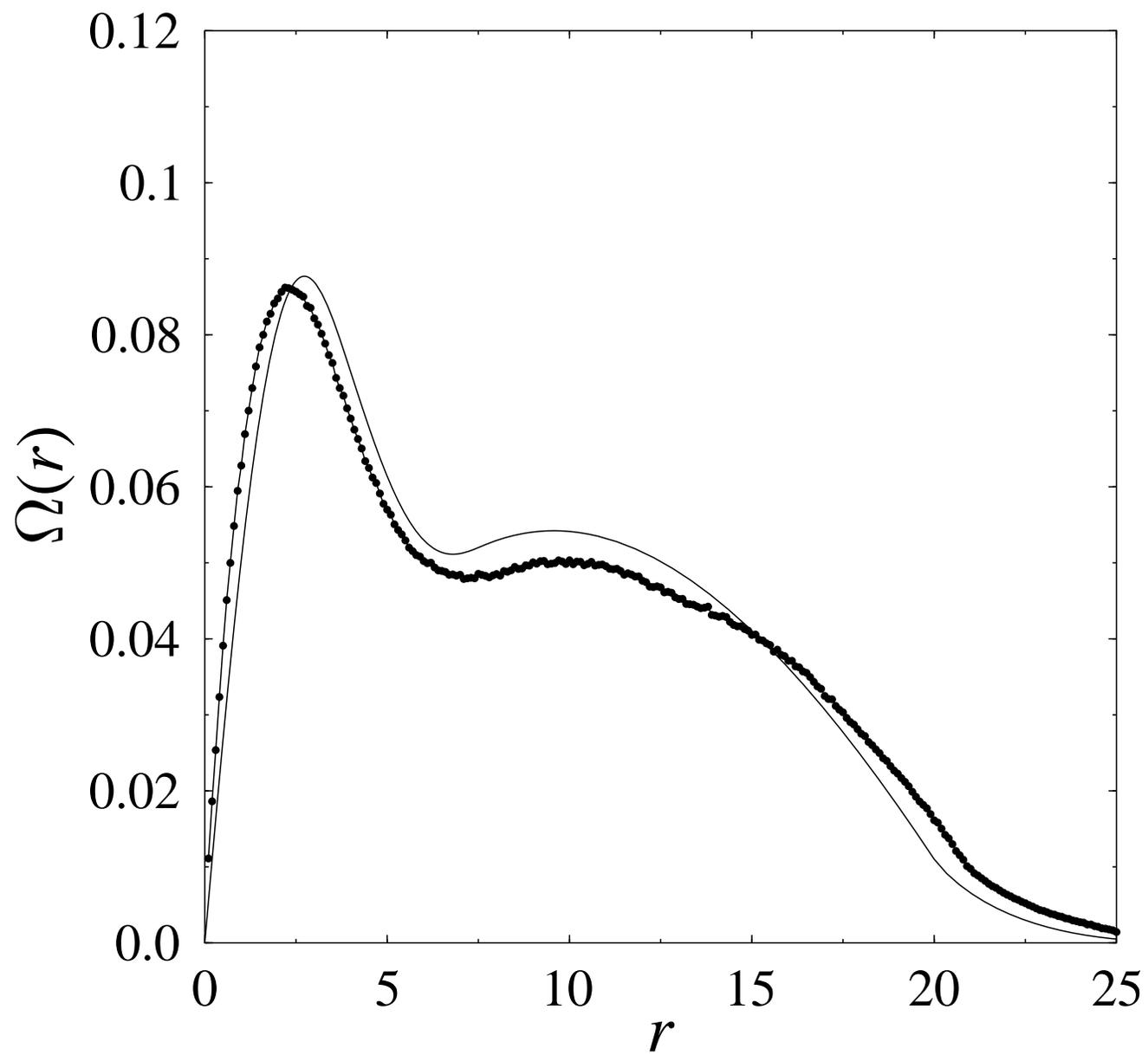
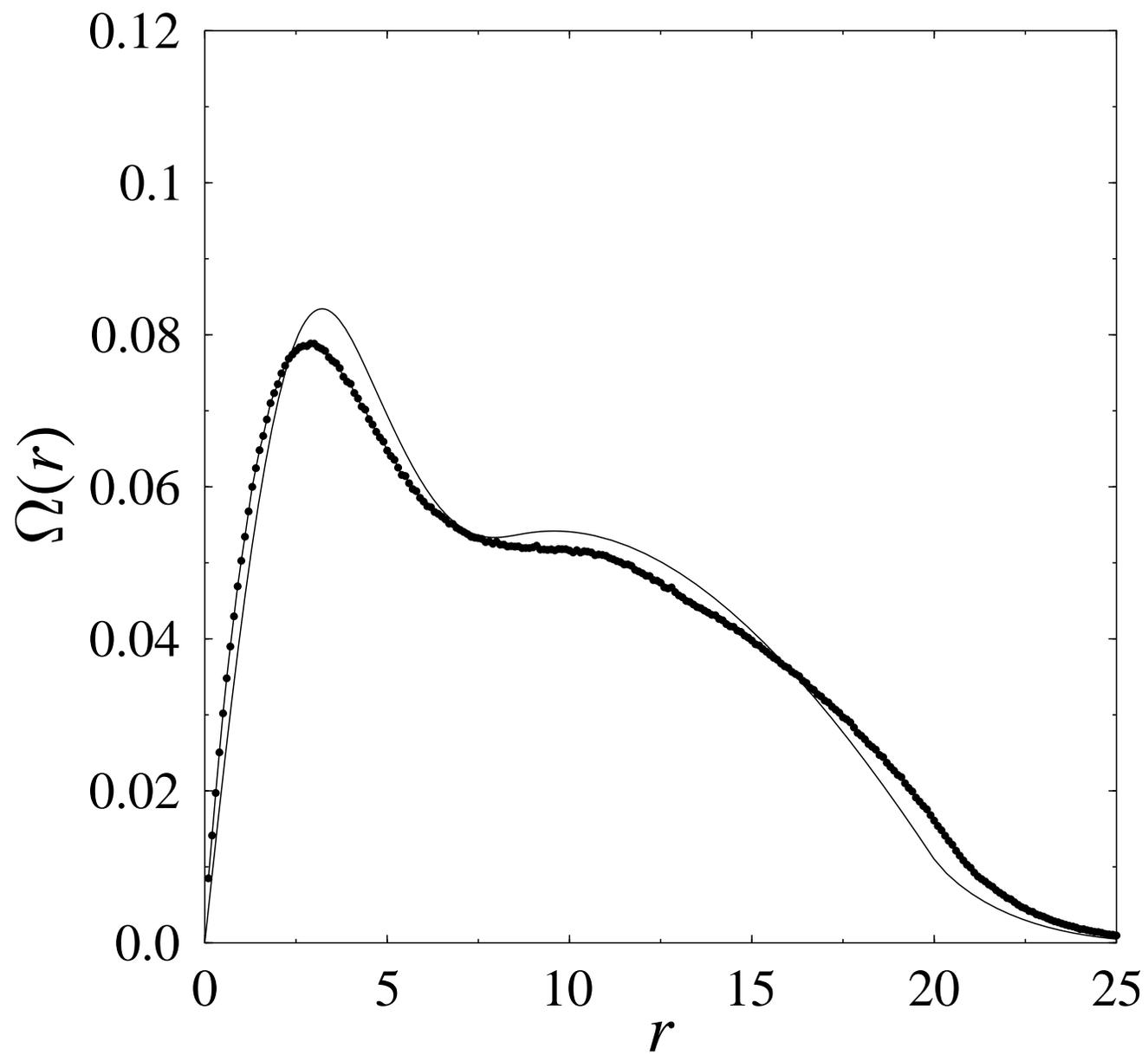
\end{document}